\documentclass[sigconf]{acmart}

\AtBeginDocument{%
	\providecommand\BibTeX{{%
			\normalfont B\kern-0.5em{\scshape i\kern-0.25em b}\kern-0.8em\TeX}}}

\copyrightyear{2025}
\acmYear{2025}
\setcopyright{cc}
\setcctype{by}
\acmConference[CHI '25]{CHI Conference on Human Factors in Computing Systems}{April 26-May 1, 2025}{Yokohama, Japan}
\acmBooktitle{CHI Conference on Human Factors in Computing Systems (CHI '25), April 26-May 1, 2025, Yokohama, Japan}\acmDOI{10.1145/3706598.3713768}
\acmISBN{979-8-4007-1394-1/25/04}

\newcommand{\hn}[1]{#1}

\usepackage{enumitem}
\usepackage{multirow}
\usepackage{subcaption}
\usepackage{amsmath}
\usepackage{soul}
\usepackage{makecell}

\makeatletter

\newenvironment{shiftedflalign*}{%
	\start@align\tw@\st@rredtrue\m@ne
	\hskip\parindent
}{%
	\endalign
}
\makeatother

\def\new{}

\begin{document}
	
	\title[You Shall Not Pass: Warning Drivers of Unsafe Overtaking Maneuvers on Country Roads]{You Shall Not Pass: Warning Drivers of Unsafe Overtaking Maneuvers on Country Roads by Predicting Safe Sight Distance} %

	\author{Adrian Bauske}
	\orcid{0009-0002-5371-8668} 
	\affiliation{%
		\institution{University of Bayreuth}
		\city{Bayreuth}
		\country{Germany}
	}
	\email{adrian.bauske@uni-bayreuth.de}
	\author{Arthur Fleig}
	\orcid{0000-0003-4987-7308} 
	\affiliation{%
		\institution{Center for Scalable Data Analytics
and Artificial Intelligence, University of Leipzig}
		\city{Leipzig}
		\country{Germany}
	}
	\email{arthur.fleig@uni-leipzig.de}

	\begin{abstract}
		Overtaking on country roads with possible opposing traffic is a dangerous maneuver and many proposed assistant systems assume car-to-car communication and sensors currently unavailable in cars.
		To overcome this limitation, we develop an assistant that uses simple in-car sensors to predict the required sight distance for safe overtaking. 
		Our models predict this from vehicle speeds, accelerations, and 3D map data. 
		In a user study with a Virtual Reality driving simulator (N=25), we compare two UI variants (monitoring-focused vs scheduling-focused). 
		The results reveal that both UIs enable more patient driving and thus increase overall driving safety.
		While the monitoring-focused UI achieves higher System Usability Score and distracts drivers less, the preferred UI depends on personal preference.
		Driving data shows predictions were off at times.
		We investigate and discuss this in a comparison of our models to actual driving behavior and identify crucial model parameters and assumptions that significantly improve model predictions. %
	\end{abstract}

	\ccsdesc[500]{Human-centered computing~Empirical studies in HCI}
	\ccsdesc[300]{Human-centered computing~User studies}
	
	\keywords{driving assistant, automotive, human-machine interaction, study}
	
	\begin{teaserfigure}
		\includegraphics[width=\textwidth]{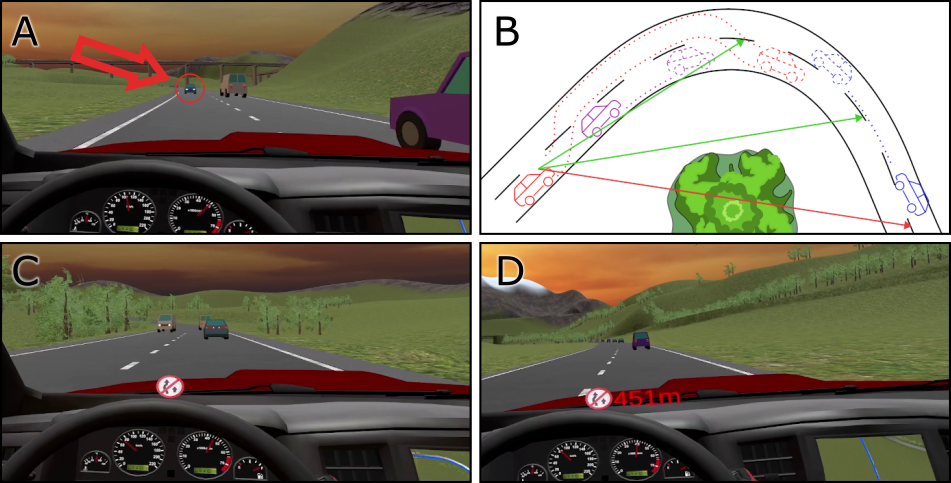}
		\caption{
			(A) Collisions with oncoming traffic when overtaking on rural roads can be fatal. 
			(B) We enhance the driver's decision-making by computing the necessary sight distances for safe overtaking maneuvers on country roads, taking into account vehicle speeds, accelerations, and 3D map data. We compare two UI variants that use these predictions:
			(C) Our monitoring-focused UI displays a warning icon on the heads-up display whenever the sight distance is insufficient for safe overtaking.
			(D) Our scheduling-focused UI additionally displays the distance (in meters) until the next overtaking opportunity.}
		\Description{This figure is divided into four parts: A, B, C and D. Part A shows a car in the process of overtaking another car on a two lane rural road. The overtaking car is roughly next to the overtaken car. On the opposing lane there is a car close to the overtaking car, forming a very dangerous overtaking maneuver. Part B shows a schematic of a curved road with a tree in the center of the curve. A typical overtaking maneuver is depicted on the road with an overtaking car, an overtaken car and an opposing car. Furthermore, sight lines to the opposing lanes are drawn, some of which the tree blocks. Part C shows the monitoring-focused UI. An icon that warns of overtaking is displayed behind the steering wheel. The car is driving on a regular country road with cars in front and cars on the opposing lane. Part D shows conceptionally the same situation as Part C, a car driving on a country road with an active overtaking warning, but the UI additionally displays the information 451m in red on the right side next to the warning icon.}
		\label{fig:teaser}
	\end{teaserfigure}

	\maketitle

	\section{Introduction}\label{sec:intro}
    Imagine driving on a two-lane country road in the car marked by the red arrow in Figure~\ref{fig:teaser}(A). 
    How to avoid such potentially fatal situations and make overtaking safer? %
	Technical assistance systems can significantly improve safety and help reaching the UN General Assembly's adopted resolution 74/299 \textit{Improving global road safety}~\cite{web:un2020:resolution-improving-global-safety} %
 to halve the number of road traffic deaths and injuries by 2030. 
    As we are transitioning towards vehicle automation, there is extensive literature on overtaking on motorways and multi-lane roads, see e.g., \cite{YU201410, Frank2021, electronics12010051}. 
    Commercially available driving assistants are able to automatically overtake on multi-lane freeways~\cite{web:mercedes-benz-youtube2023}, %
    where there is no oncoming traffic.

    However, UNECE statistics~\cite{for2022statistics} show that road traffic accidents are far more common on country roads than on motorways. 
    These roads are often narrow, with blind curves and junctions, offering few safe overtaking spots. 
    Traffic is diverse, including cyclists (vulnerable road users~\cite{vulnerableRoadUsers}) and slow farm vehicles. 
    Some drivers feel "compelled" to overtake and frequently search for overtaking opportunities~\cite{KINNEAR2015221}. 
    As many dashcam videos on platforms like YouTube %
    show, some drivers take greater risks with shorter sight distances, endangering themselves and others.

	Yet, to the best of our knowledge, no commercially available assistance system for overtaking on \textit{rural} roads exists. 
	This is because a big challenge is sensing or confidently predicting oncoming traffic in due time. 
	Sensors installed in cars today are unsuitable for this task, e.g., because of their limited range. 
	As an alternative, many concepts require vehicle-to-vehicle (V2V) and/or vehicle-to-interface (V2I) communication~\cite{Han-Shue2006,Toledo2009}, which is not yet (widely) available. 
	
	In this paper, we develop an assistant for country roads that operates without these concepts and works within the range of current in-car sensors.
    Our assistant does not overtake autonomously, and instead enhances the driver's decision-making process before initiating any overtaking maneuver. 
    Our main idea is to predict the required sight distance for safe overtaking and warn drivers of dangerous overtaking maneuvers, such as in Figure~\ref{fig:teaser}(A).
	We base our prediction on vehicle speeds, accelerations, and 3D map data (Figure~\ref{fig:teaser}(B)).
	We introduce two UI variants: the monitoring-focused UI, which shows a warning icon when sight distance is insufficient (Figure~\ref{fig:teaser}(C)), and the scheduling-focused UI, which provides additional information like the distance to the next overtaking opportunity.
	We refer to these instances as opportunities as our system identifies situations where visibility is insufficient.
	If no warning is displayed, %
	it is still the driver's responsibility to check for oncoming traffic and execute the overtaking maneuver. 
	
	We assess the potential of our assistant and the two UIs in a user study ($N=25$) on a Virtual Reality (VR) driving simulator. 
	The following questions, for which we derive hypotheses in Section~\ref{subsec:hypotheses}, guide our research:
	\begin{enumerate}[label=\textbf{RQ.\arabic*}]
		\item\label{item:rq-acceptance} %
			How do drivers perceive and accept our assistant (in terms of workload and usability) across the UIs?%
		\item\label{item:rq-safety} Which UI helps drivers (more) to avoid dangerous overtaking maneuvers? 
		\item\label{item:rq-models} %
			How accurate are the sight distance prediction models used by our assistant? %
	\end{enumerate}

	In summary, we \textbf{contribute} a driving assistant with two UI variants, enabled by new models for sight distance prediction.
    We evaluate the assistant and compare the UIs and models in a user study, and extensively discuss the results. 
	We open-source the sight distance prediction models at \url{https://github.com/abauske/YouShallNotPass}
	Moreover, with TOWARDS~\cite{dataset:towards}, we make the driving data from the user study on the VR driving simulator -- with 1240 minutes of driving (1363 minutes simulation time including breaks) and 576 completed overtaking maneuvers -- freely available. %

	\section{Background and Related Work}\label{ref:relwork}
	\subsection{Available Sensors and Data, and Constraints}\label{subsec:sensors-available}
	Ahangar et al.~\cite{Ahangar2021} provide a comprehensive list of currently available in-car sensors and their capabilities. 
	This includes 
	ultrasonic sensors (distant measurement to objects up to 3m away), radio detection and ranging (RADAR) (presence detection of and distance measurement to objects in a range of 5 to 200m, also usable for determining relative speed), light detection and ranging (LiDAR) (obstacle detection and mapping of environment up to 200m), cameras (environment surveillance, up to 250m), and GPS (position and speed measurement for navigation). 
	Currently available 3D map data includes road types, radius and slope data of curves, number of lanes and their width, road quality and hazards, restrictions such as speed limits and overtaking bans, crosswalks, and even traffic information~\cite{trafficData, DynamicPassPrediction}.

	We intentionally limit ourselves to this existing information and %
	\textit{exclude} the following: 
	\begin{enumerate}
		\item \textit{Long-Range Opposite Lane Detection:} 
		Safe overtaking relies on reliable long-range detection of oncoming vehicles, but current in-vehicle sensors (range: 250m) fall short of the required range (up to 900m)~\cite{NCHRP_NAP23278}.
		\label{enum:constraints:range}
		
		\item \textit{Vehicle Communication:} Vehicle-to-Vehicle (V2V), Vehicle-to-Infrastructure (V2I), and Vehicle-to-Everything (V2X) communication concepts exist, but they require widespread vehicle adoption and face technical challenges. 
		
		\item \textit{Aggregated Data from Connected Vehicles:} Using preceding connected vehicles to extend sensor range requires reliable communication and is not feasible under current conditions.
	\end{enumerate}

	\subsection{Existing Overtaking Assistants}\label{subsec:existing-assis}
	
	The literature on overtaking assistants and autonomous passing maneuvers is extensive.
	
	Toledo-Moreo et al.~\cite{Toledo2009} propose a cooperative overtaking prediction system. 
	Similar to our work, it is based on vehicle kinematics and road geometry. 
	However, this system communicates predictive information via mobile networks, including trajectory predictions of oncoming vehicles and relies heavily on vehicle-to-vehicle communication, which we exclude.
	Similarly, in~\cite{Han-Shue2006} each vehicle identifies potential collisions based on its own measurements and exchanges information regarding their locations, intentions, and other relevant details with other vehicles and infrastructure.

	Hassein et al.~\cite{Hassein2019} develop a Passing Collision Warning System for trucks, using cameras and RADAR to detect oncoming vehicles and warn drivers of dangers during overtaking. 
	While promising, it requires sensor ranges beyond what is currently available, unlike our assistant, which operates within existing sensor ranges.

	Commercial vehicles offer automated overtaking assistants combining lane-keeping and lane change systems. 
	While some require user action to initiate the maneuver (turn signal activation~\cite{web:tesla:autopilot}, %
    looking in the side mirror~\cite{web:usatoday-bmw}), %
    others do not~\cite{web:mercedes-benz-scottsdale}. %
	Yet, to the best of our knowledge, these systems are designed for multi-lane freeways and do not assist with overtaking on roads with opposing traffic.
	Moreover, the driver must be able to take over control within a few seconds.

Yet, Hegeman et al.~\cite{Hegeman_Brookhuis_Hoogendoorn_2005} claim that an active assistant could handle most overtaking-related tasks. 
While they do not develop a complete system due to identified sensor limitations, they provide a comprehensive analysis of overtaking maneuvers.
We draw from their rich descriptions of the many important aspects of an overtaking assistant.

Finally, Loewenau et al.~\cite{DynamicPassPrediction} present a passive system that uses vehicle dynamics and navigation data to identify unsuitable road segments for overtaking. 
While it does not rely on communication networks or sensors not yet available in cars, it does not consider oncoming traffic.
In contrast to our system, it does not rely on sight distance and lacks a user study. 
Yet, their use of navigation data, particularly road curvature, inspired elements of our approach.

\subsection{User Interfaces in Overtaking Assistants}\label{subsec:uis-in-existing-assistants}

Concerning the interfaces, we %
classify overtaking assistants into two groups. 
The first group are systems that overtake autonomously. 
Research in this setting has explored various aspects of Human-Machine Interfaces (HMI). 

Fank et al.~\cite{Frank2021} examined different HMI designs for cooperative overtaking between trucks on highways.
In these situations, speed differences are low, and a larger amount of information can be displayed, such as animated traffic flow, speed limits, active assistants, and more.
Yet, they identified a narrow margin between providing too much and too little information for drivers to accept the assistant. 
To address this, we investigate both a minimalistic \textit{monitoring-focused} UI focusing on essential information, and a more detailed \textit{scheduling-focused UI} that provides extended information.

Walch et al.~\cite{Walch2018,Walch2019} consider overtaking on country roads and thus a case closer to ours.
Their assistant overtakes autonomously once the driver approves the maneuver.
In a driving simulator, they compared two user approval and cancel methods. %
While participants preferred the assistant over manual overtaking, they tended to overlook rear traffic in more complex situations. 
This indicates a need for support in the critical decision-making phase.

In contrast, our assistant emphasizes enhancing the driver's decision-making process before initiating any overtaking maneuver and falls into the second group of \textit{passive} assistants, whose 
main goal is to convey information to the driver regarding overtaking safety. 
User interaction happens by driving the car.
While Hassein et al.~\cite{Hassein2019} from Section~\ref{subsec:existing-assis} focus on the technical side and suggest a (possibly color-coded) "(Not) Safe to pass" text message at an unspecified location, Hegeman et al.~\cite{Hegeman2007} assume world knowledge and design a traffic light GUI placed to the right of the rear mirror. 
It lights up green or red, depending on whether overtaking is currently safe or not. 
Three seconds before changing its color, it starts blinking, allowing for some scheduling. 
We draw from the idea to enable \textit{scheduling} and \textit{monitoring}. 

Furthermore, gamified user interfaces have been explored, e.g., by Steinberger et al.~\cite{Steinberger2016}, suggesting higher user acceptance but also increased visual distraction. 
The reported increase in long eye glances, which is particularly dangerous on curvy country roads, made us hesitate to adopt this approach to our case.

\subsection{Passing Sight Distance \new{and the Point-of-no-return}}\label{subsec:psd}

Our assistant relies on a sight distance prediction model developed in this work. 
The required sight distance for safe overtaking, known as Passing Sight Distance (PSD), is widely studied for road design~\cite{Glennon1988NEWAI, harwood1989passing} and road marking~\cite{Valkenburg, Saito1984}. 
Haq et al.~\cite{HAQ2022255} use Glennon’s model~\cite{Glennon1988NEWAI} for computing the PSD for overtaking truck platoons.
However, all these models are designed for static evaluation, assuming a constant speed difference between overtaking and overtaken vehicle, neglecting acceleration.
Lieberman~\cite{Lieberman1982} introduces a model that includes acceleration, but it has been criticized for overestimating PSD by including the distance to a point-of-no-return, until which an overtaking maneuver can be aborted~\cite{NCHRP_NAP23278}. 
Our model incorporates aborting a maneuver, focusing on the PSD required at the point-of-no-return.

\new{There are several definitions of the point-of-no-return in literature: when vehicles are head to head~\cite{HASSAN1996453}, when their centers align~\cite{Saito1984}, or when the overtaken vehicle's rear bumper aligns with the overtaking vehicle's center~\cite{Valkenburg}.
Given the proximity of these definitions and no clear consensus in the literature, we follow Saito~\cite{Saito1984}, considering the maneuver committed once the vehicle centers align.}

Raj et al.~\cite{Abhishek2023} extend Glennon’s model by considering road gradient and tire friction, factors that influence acceleration. 
However, the lacking mathematical description of the longitudinal acceleration and the model's inability to adapt to dynamic traffic conditions greatly limit its applicability. 
Our assistant addresses these gaps by measuring real-time speed differences and explicitly including longitudinal acceleration.

In addition to calculating PSD at the point-of-no-return, our assistant needs to predict this critical point itself, rendering accurate acceleration models essential. 
We thus provide three such models with increasing complexity to enhance prediction accuracy in Section~\ref{subsec:acc-models} and compare them.

\section{The Overtaking Warning Assistant}\label{sec:assistant}
We introduce the assistant's core concept and design considerations in Section~\ref{subsec:core-concept}, explain our design rationale behind the developed UIs in Section~\ref{subsec:uis}, and present our underlying prediction model 
for the safe sight distance 
in Section~\ref{subsec:prediction-model}.

\subsection{Design Considerations and Core Concept}\label{subsec:core-concept}
The sensor constraints imposed in Section~\ref{subsec:sensors-available} prevent communication with or timely sensor-based recognition of oncoming traffic on country roads. 
Hence, we cannot implement safe autonomous overtaking. 
Instead, our assistant warns of dangerous overtaking maneuvers so that drivers do not initiate them in the first place.

Before designing the assistant, we have analyzed the driving task and established the following key 
	design considerations
	to guide the development, subject to the technical possibilities and limitations.
\begin{enumerate}[label=\textbf{C.\arabic*}]
	\item\label{item:design-goal1} Warnings should appear if and only if they are appropriate. 
	\item\label{item:design-goal2} The drivers should not need to 
	take their view from the (potentially curvy) road to use the assistant. 
	\item\label{item:design-goal3} Warnings and other information should be easy to learn and understand.  %
	\item\label{item:design-goal4} The interface should not distract the user, ideally taking off cognitive load. 
\end{enumerate}

Addressing~\ref{item:design-goal1} is crucial for user acceptance but challenging because it intertwines both human-centered elements, such as displaying information, and machine-centered elements, such as the underlying prediction model. 
Adhering to the sensor limitations and with~\ref{item:design-goal1} in mind, the \textbf{core concept} of the assistant is as follows.

The assistant is triggered whenever a vehicle in front is within sensor-range.
It then models and computes on a moment-to-moment basis (once per second) the point-of-no-return \new{(when vehicle centers align)} and the passing sight distance at the point-of-no-return.
This requires a carefully designed model that \textit{dynamically} predicts the %
passing sight distance and the point-of-no-return, adapting to traffic and user interaction. 
Since, to the best of our knowledge, no such model exists, we develop it in Section~\ref{subsec:prediction-model}, closely adhering to existing literature for individual components. 
As long as the PSD exceeds the available sight distance, e.g., due to curves or obstacles such as trees, our assistant displays a warning (and additional information depending on the UI). %
The calculations take into account intersections (where one should not overtake~\cite{Hegeman_Brookhuis_Hoogendoorn_2005}) and whether there is enough space between two leading vehicles within sensor-range.
Ideally, the driver recognizes that overtaking is too risky and follows the vehicle in front instead. 
If no warning is displayed, the sight distance is sufficient; however, the driver must still check for oncoming traffic, ensure their view is not obstructed by lead vehicles or other dynamic objects not detectable from map data, and execute the overtaking maneuver.

\subsection{UI Design Rationale: Monitoring and Scheduling}\label{subsec:uis}
\begin{figure}[ht]
    \centering
    \begin{subfigure}[ht]{\linewidth}
    \includegraphics[width=1\textwidth]{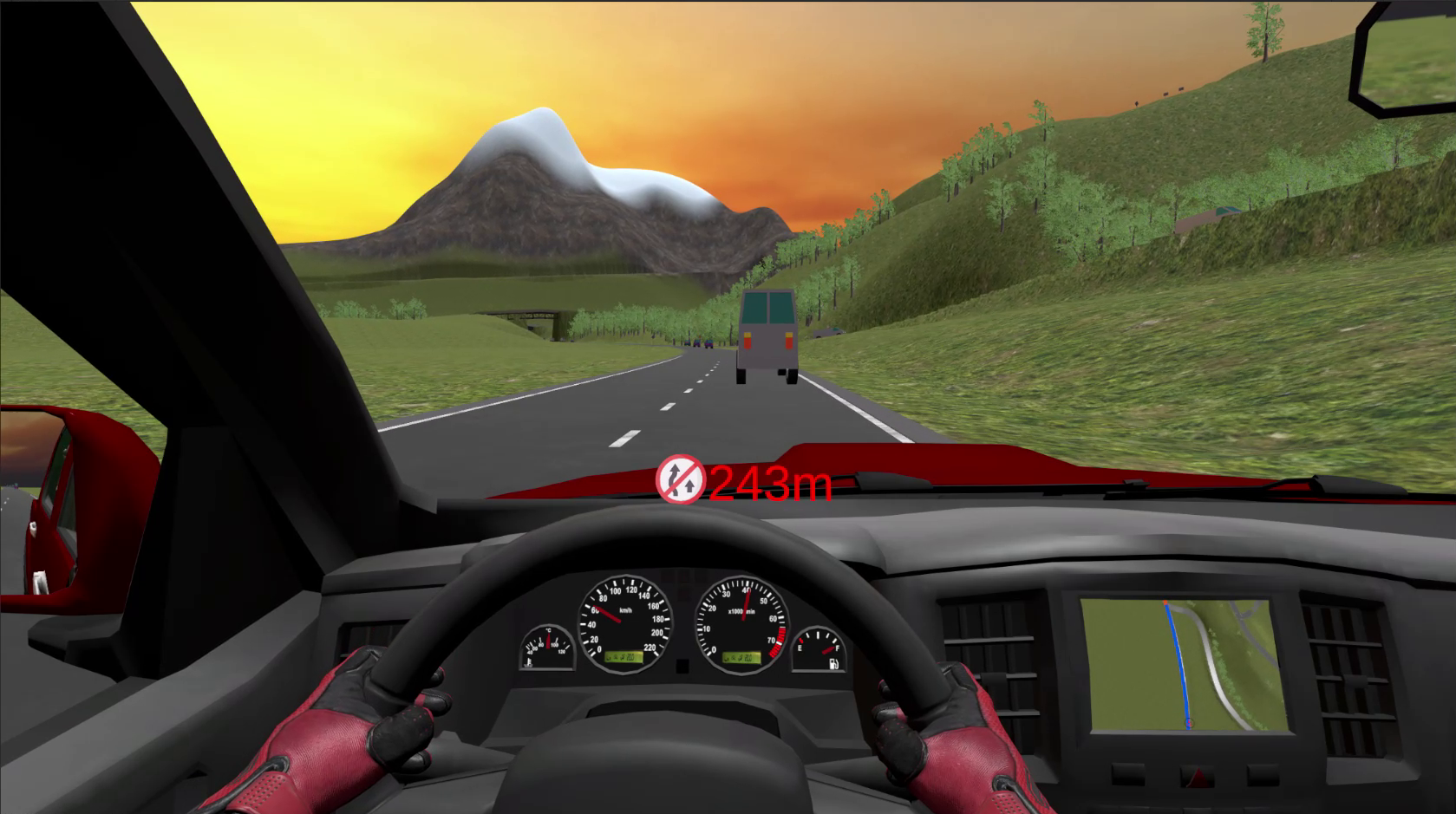}
    \end{subfigure}
    \begin{subfigure}[ht]{\linewidth}
    \includegraphics[width=1\textwidth]{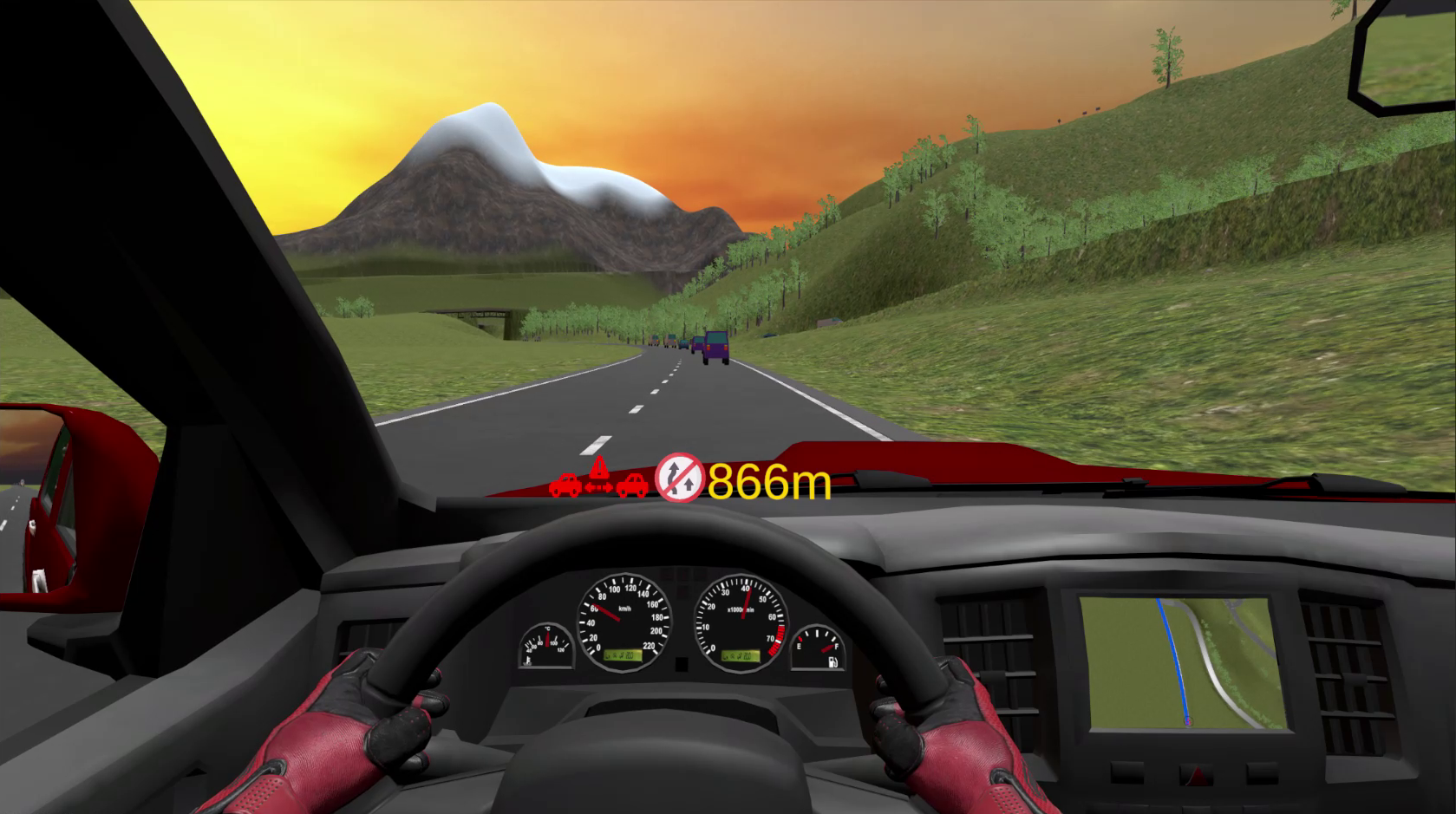}
    \end{subfigure}
    \caption{HUD with the scheduling-focused UI. 
    (Top) Overtaking now is not safe (warning icon). 
    The next overtaking opportunity (in red due to potential oncoming traffic) is in 243m. 
    (Bottom) Overtaking now is not safe. %
    There is not enough room between two vehicles ahead (red warning icon to the left). 
    The next overtaking opportunity (in yellow due to a second lane opening) is in 866m. \label{fig:ui-variants}}
    \Description{This figure is split into a top and a bottom figure. Both show a car driving on a two lane rural road with one car in front in the top figure and multiple cars in front in the bottom figure. The main deviation between the two figures is the HUD. In the top figure there is an icon signaling to not overtake in the middle and to its right it says in red color: 243m. In the bottom figure there is the same "Do not overtake" sign in the middle and to its right it says in yellow color: 866m. Additionally, there is an icon on the left side of the HUD signaling there is not enough room between two vehicles in front.}
\end{figure}

To address~\ref{item:design-goal2}, we chose a heads-up display (HUD) because it allows drivers to monitor information without losing sight of the road and it does not interfere with other instruments \new{-- speedometer and rpm gauge are displayed on a standard dashboard interface below the steering wheel.}
We positioned the HUD just above the steering wheel (Figure~\ref{fig:ui-variants}), minimizing the field-of-view limitations of the VR headset used in our study.
\new{
In this prototype, the HUD was reserved for the assistant’s functionality and emphasized the new, while the classic dashboard conveyed a sense of familiarity.}%

\new{
Our UI designs draw inspiration from Hassein's~\cite{Hassein2019} suggestion of a (possibly color-coded) "(Not) Safe to pass" text message, Hegeman's~\cite{Hegeman2007} traffic light GUI that allows for some planning, and Fank's~\cite{Frank2021} identified narrow margin between providing too much and too little information in the context of cooperative overtaking between trucks on highways. 
We held formative discussions with (senior) HCI researchers, who did not participate in the user study. 
Their feedback and our own test-driving helped in developing two UIs.}
Our minimalistic \textbf{monitoring-focused UI} (Figure~\ref{fig:teaser}(C)) displays a warning icon if the sight distance is insufficient and only has two operational states: 'warning' or 'no warning'.
Our \textbf{scheduling-focused UI} (Figure~\ref{fig:ui-variants}) provides more details and highlights overtaking opportunities. %
The distance to the next overtaking opportunity is shown on the right. 
If red (Figure~\ref{fig:ui-variants} (top)), it indicates the distance to the next point with sufficient sight distance, though overtaking is still risky due to potential oncoming traffic. 
If yellow (Figure~\ref{fig:ui-variants} (bottom)), it indicates a second lane opening, a safer spot with no opposing traffic. 
Additionally, if there is insufficient space between two vehicles ahead within sensor-range, a red warning icon %
appears to the left of the main warning icon. 
The monitoring-focused UI just displays the "no overtaking" warning in this case.

To address~\ref{item:design-goal3}, we chose distinct and intuitive warning icons. 
\new{The "no overtaking" warning icon conveys overtaking-related guidance, while not taking up much space.}
We specifically did not use the "overtaking prohibited" sign to not confuse drivers into thinking the assistant recognizes overtaking bans. 
\new{None challenged the icon in formative discussions and test drives and we kept it for our initial design.}
Being the most important information, it is placed in the center of the HUD, always in the (peripheral) view to allow for quick reaction. 
\new{Since the dashboard interface already displayed the speed, the assistant’s warning could occupy a prominent position without competing with standard HUD information.}
The color-coded distance to the next overtaking opportunity represents the risk of overtaking.
\new{When sight distance is sufficient, nothing is displayed on the HUD. 
Since we did not implement a deactivation function, 
the absence of a warning had a single, clear cause. 
This design ensures warnings appear only when necessary.
}

Regarding~\ref{item:design-goal4}, \new{we deliberately varied the level of detail in the two UIs to explore the trade-offs between simplicity and additional context.} 
With both UIs, drivers need to focus on oncoming traffic only once the warning disappears and they wish to overtake. 
However, the scheduling-focused UI \new{might allow drivers to better align their decisions with traffic conditions.} 
Whether this \new{additional} information is too much is a question we investigate with this initial design.

\subsection{Predicting Safe Sight Distance}%
\label{subsec:prediction-model}

\subsubsection{The Overtaking Maneuver}\label{subsec:overtaking-maneuver}
\begin{figure}[htb]
    \includegraphics[width=\linewidth]{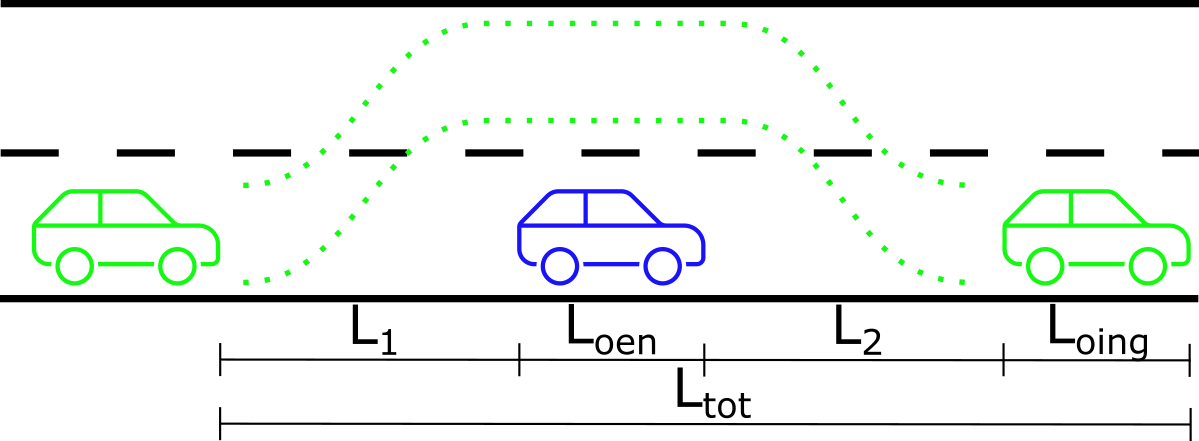}
    \caption{Static distances in a typical overtaking maneuver. 
    $L_1$ and $L_2$ are the distances to the vehicle to be overtaken in phase (i) and after phase (iii), respectively.
    $L_{oing}$ and $L_{oen}$ are the lengths of the overtaking car %
    and the vehicle to be overtaken, respectively. %
    }
    \label{fig:overtaking-maneuver-simple}
    \Description{This figure shows a schematic of a straight road with three cars on the right lane with the first and last car having the same green color and the middle one being blue. There is a dotted green line from the front of the car that is last in line ("last car") to the end of the car that is first in line ("first car"), transitioning to the opposing lane passing the blue car and transitioning back to the right lane. Furthermore, distance definitions are depicted: L_1 is the distance of the last car's front to the blue car's back. L_oen is the length of the blue car. L_2 is the distance from the blue car's front to the first car's back. L_oing is the length of the first car. L_tot is the sum of the previously mentioned lengths, which corresponds to the distance of the last car's front to the first car's back.}
\end{figure}

According to~\cite{Petrov2014}, overtaking consists of three phases: (i) changing lanes, (ii) driving alongside the other vehicle, and (iii) returning to the original lane, as shown in Figure~\ref{fig:overtaking-maneuver-simple}.
For the computation of required sight distances, we model the total distance traveled by the overtaking ego-car. 
While models for the lateral movement in phases (i) and (iii) exist, cf.~\cite{Hassein2019}, we consider only the longitudinal distance to greatly simplify computations. 
If the overtaking maneuver was completed within 10m longitudinal distance, e.g., overtaking a stationary object, the introduced error would be as high as 5\%. 
However, %
the error is <0.1\% for 60m longitudinal distance %
and <0.01\% for 100m, which is a typical length for overtaking at 65km/h~\cite{Abhishek2023}.
In summary, the introduced error is negligible for the high speeds we consider.
The total \textit{static} distance traveled depicted in Figure~\ref{fig:overtaking-maneuver-simple} is
\begin{equation}\label{eq:ltot}
L_{tot} = L_1 + L_{oen} + L_2 + L_{oing}.
\end{equation}
Adding the distance covered by the overtaken car, $d_{oen}$, gives the total distance: %
\begin{equation}
	d_{tot} = L_{tot} + d_{oen}. 
\end{equation}

\subsubsection{Modeling Acceleration}\label{subsec:acc-models}
The total distance traveled by the ego-car %
crucially depends on its acceleration. 
Mathematically, if we denote the acceleration by $a(t)$ as a function of time $t$, then integrating it once yields the current velocity~$v(t)$ and integrating it twice yields the distance traveled up to time $t$, which we denote by~$s(t)$. 

For our assistant, we consider three acceleration models of increasing complexity:
\begin{itemize}
\item 
\textit{(Constant)} The simplest model assumes constant acceleration~$a_{const}$, but takes into account the road slope:
\begin{equation}\label{eq:acc-const}
a = a_{const} - a_G,
\end{equation}
where $a_G$ is the gravitational constant $g=9.81$ multiplied by $G$, the road slope in percent.
We set $a_{const}$ to 3 m/s$^2$. 
This model's main advantage is its simplicity, which reduces computational load.
\item 
\textit{(LDM)} The intermediate model is a linear-decay-model~\cite{acceleration_models}. It also takes into account the road slope, but the constant acceleration~$a_{max}$ linearly decreases with speed~$v$ up to a certain speed~$v_e$:
\begin{equation}\label{eq:acc-ldm}
a = a_{max} - a_G - \frac{a_{max}}{v_e} \cdot v.
\end{equation}
We set $v_e=41.85 m/s$ and $a_{max}=7.8 m/s^2$.
\item 
\textit{(Dynamic)} The third model is based on forces~\cite{Rakha2002,acceleration_models}. 
Here, the resistant force~$F_R$ is deducted from the driving force~$F_A$ and the result is divided by the vehicle mass~$m$ (2300 kg) to yield the acceleration:
\begin{equation}\label{eq:acc-dyn}
a = \frac{F_A - F_R}{m}.
\end{equation}
Its complexity comes from modeling the forces.
Resistant forces include air resistance, wheel friction, and road incline.
Driving forces depend on, e.g., motor power, efficiency, and friction.         
Detailed formulas and parameters are in Appendix~\ref{app:dyn-acc}.
The main advantages are accurate vehicle behavior prediction and flexibility in estimating acceleration for various vehicle sizes and terrains using readily available input parameters.
\end{itemize}
Taking the acceleration as given by the models means drivers floor the gas pedal. 
As we assume this is not always true, we multiply each model's acceleration~$a$ by a coefficient~$\lambda$, which we set heuristically to 0.8.

These models predict the total distance traveled~$s(t)$ during a potential overtaking maneuver, which is compared to available sight distance to assess safety. 
For the "Constant" model, an analytic formula for~$s(t)$ exists. 
For the other two models, we compute~$s(t)$ numerically using the Forward Euler scheme~\cite{butcher2016numerical}.
Figure~\ref{fig:acc-model-comparison} illustrates the three models (with chosen parameters as detailed in Appendix~\ref{app:dyn-acc}) and VR test drive data, with the \textit{Dynamic} model providing the best fit.

\begin{figure}[ht]
	\includegraphics[width=\columnwidth,trim={0cm 0 0.1cm 0},clip]{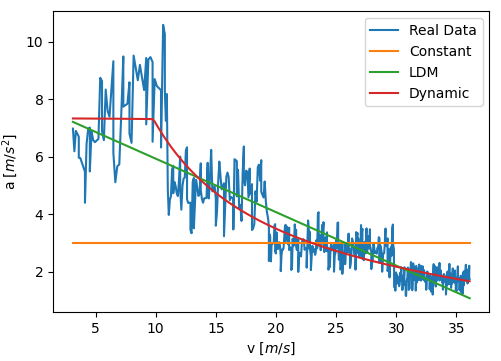}
	\caption{Comparison of the three acceleration models to measured data from the driving simulator ("real data"), in dependence of velocity. The clearly visible steps in the measured data stem from switching gears. \label{fig:acc-model-comparison}}
    \Description{This figure shows a graph with the acceleration on the y axis and velocity on the x axis. There are four different plotted curves. The blue curve is marked as "Real Data" and is noisy. It starts at about 8 m/s^2 at about 5 m/s and reduces over three steps to about 2 m/s^2 at 35 m/s. From one step to the next the value is noisy but relatively constant. The red curve is marked as "Dynamic" and is constant at low velocities of up to 10 mm/s, at the same height as the blue curve. At 10 m/s it starts declining smoothly to the same end values as the blue curve. The green ("LDM") curve shows a linearly decreasing fit of the blue curve. The yellow curve ("Constant") is just a horizontal line at 3 m/s^2.}
\end{figure}

\subsubsection{Modeling Overtaking Behavior}\label{subsec:overtaking-beh-model}
Our assistant considers a maximum overtaking speed, denoted by $v_{oing}^{max}$, regardless of the acceleration model.
To follow road regulations, $v_{oing}^{max}$ is based on the current speed limit (60-100 km/h).
To account for deviations reported in \cite{NAP22048,Mannering2007EffectsOI,HAGLUND200039}, a 5\% tolerance is added on top, capping $v_{oing}^{max}$ at 105km/h.

When returning to one's own lane in phase (iii) of the overtaking maneuver (see Section~\ref{subsec:overtaking-maneuver}), braking might be necessary, e.g., because of an upcoming speed limit or a vehicle ahead. 
According to~\cite{Wilson1941}, a comfortable deceleration is less than 2.61m/s$^2$, while a deceleration above 4.24m/s$^2$ is considered critical. 
To allow the driver to decelerate effectively but not alarmingly, our assistant assumes a deceleration of 3.3m/s$^2$, should it be necessary. 

Furthermore, while overtaking several vehicles in a queue happens in reality~\cite{Hegeman_Brookhuis_Hoogendoorn_2005}, our assistant assumes only one car is overtaken at a time. 
In theory, a queue could be treated as one long vehicle (to the extent of the sensor ranges). 
While this is easy to implement, the resulting sight distances are impractical and only emphasize the danger of this maneuver. 

Aborting an overtaking maneuver and returning behind the vehicle with sufficient safety distance 
is possible in phase (i), and in phase (ii) up to the point-of-no-return \new{(when the vehicle centers align, see Section~\ref{subsec:psd})}.

\subsubsection{Modeling Other Vehicles}
\begin{figure}[ht]
	\includegraphics[width=\linewidth,trim={0cm 0 0cm 0},clip]{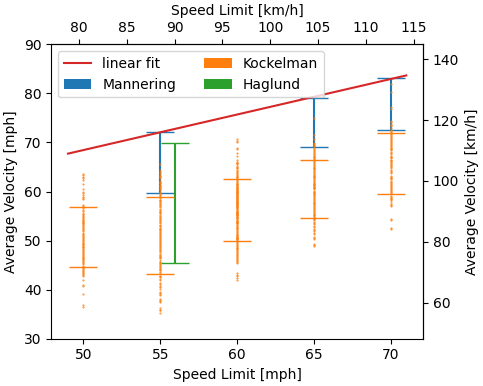}
	\caption{Typical speeds traveled at various speed limits as reported in the works by Kockelman~\cite{NAP22048}, Mannering~\cite{Mannering2007EffectsOI}, and Haglund et al.~\cite{HAGLUND200039}. Our linear model (red) covers the upper end to be on the safe side and anticipate a higher speed rather than a lower one. \label{fig:speed-limit-and-actual-speed}}
    \Description{This figure shows boxplots at various speed limits (x axis) showing the observed velocities (y axis) at given limits. There is data from three different authors: Mannering (blue and higher than Kockelman's values), Kockelman (yellow and lower than Mannering's values) and one plot by Haglund at 90 km/h speed limit spanning almost the full range of Mannering and Kockelman. Additionally, there is a linear fit of the top-most points of all other plots in red.}
\end{figure}

Since we cannot recognize oncoming traffic in time, our assistant always assumes its presence for safety reasons. 
In deciding whether the available sight distance is sufficient, we thus need to anticipate their speed. 
To this end, we build on the works by Kockelman~\cite{NAP22048}, Mannering~\cite{Mannering2007EffectsOI}, and Haglund et al.~\cite{HAGLUND200039}, who study the average driving speed for given speed limits.
This relationship is shown in Figure~\ref{fig:speed-limit-and-actual-speed}.
We are particularly interested in the expected maximum speeds of the opposing traffic, which we denote by $v_{opp}^{max}$. 
We estimate these using a linear model (red line in Figure~\ref{fig:speed-limit-and-actual-speed}).
In mathematical terms, denoting by $v_{opp}^{lim}$ [km/h] the allowed speed limit on the opposing lane (which we infer from map data), $v_{opp}^{max}$~[m/s] is calculated via
\begin{equation}\label{eq:vmaxopp}
\hn{v^{max}_{opp}} = \frac{0.725 \cdot \hn{v^{lim}_{opp}} + 51.801}{3.6}.
\end{equation}

Our assistant also considers the current speed of the vehicle to be overtaken with an update frequency of 1Hz.

\subsubsection{Calculating the Required Sight Distance $d_{s,min}$}\label{subsec:calculating-safety-distances}
To ensure safe overtaking, we compute the total required sight distance, which is composed of several safety distances. 

Two of them, $L_1$ and $L_2$, are illustrated in Figure~\ref{fig:overtaking-maneuver-simple}.
These distances are often expressed in seconds rather than units of distance to account for speed. 
When we specify \ul{safety distance in seconds, we use the superscript $^s$}, e.g., $L_1^s$.
For $L_1^s$, we assume 1s, to maintain a safe distance while not prolonging the overtaking maneuver. 
This is the lower end of a possibly user-configurable parameter. 
For $L_2^s$, according to~\cite{Hegeman_Brookhuis_Hoogendoorn_2005}, 80\% of overtaking maneuvers are below the recommended 2s, and 15\% are below 1s. 
Our assistant assumes $L_2^s=1s$, which is on the lower end of a possibly user-adjustable parameter. 

\begin{figure}[ht]
    \includegraphics[width=\linewidth]{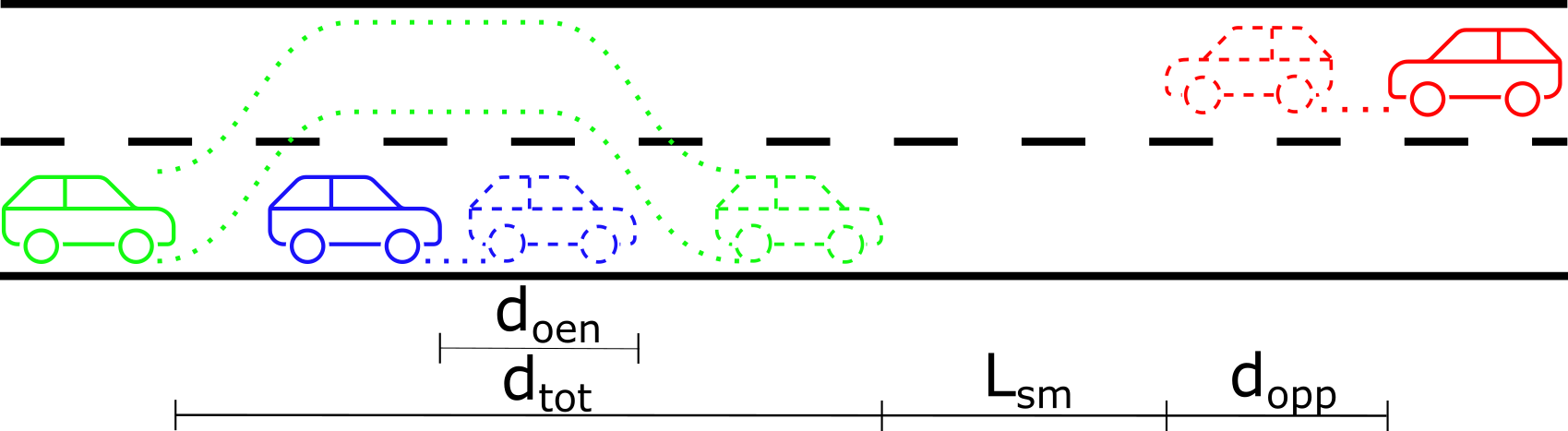}
    \caption{Safety distances when overtaking with oncoming traffic. $L_{sm}$ denotes the static distance between the overtaking vehicle (which traveled a distance of $d_{tot}$ during the maneuver) and the oncoming vehicle (which traveled a distance of $d_{opp}$ during the maneuver) after the overtaking is finished. \label{fig:safety-distance-opposing}}
    \Description{This figure shows a schematic of a straight road with cars driving on the right lane, where one car overtakes another car (dotted lines resemble their paths). There is a car on the opposing lane. Various distances are illustrated. The distance covered by the opposing car is denoted as d_opp. The distance covered by the overtaken car is denoted as d_oen. The distance between the overtaking car's front after the maneuver and the opposing car's front after the maneuver is L_sm. The distance covered by the overtaking car is d_tot.}
\end{figure}
In addition, we need to keep a safety distance from opposing vehicles, denoted by $L_{sm}$ in Figure~\ref{fig:safety-distance-opposing}. 
$L_{sm}^s$ is set to 1.5s, as suggested by~\cite{VanDerHorst1994}.
It could also be a user-configurable parameter based on recommendations in~\cite{Hassein2019} (2s) and~\cite{Hegeman_Brookhuis_Hoogendoorn_2005}.

\begin{figure}[ht]
\includegraphics[width=\linewidth]{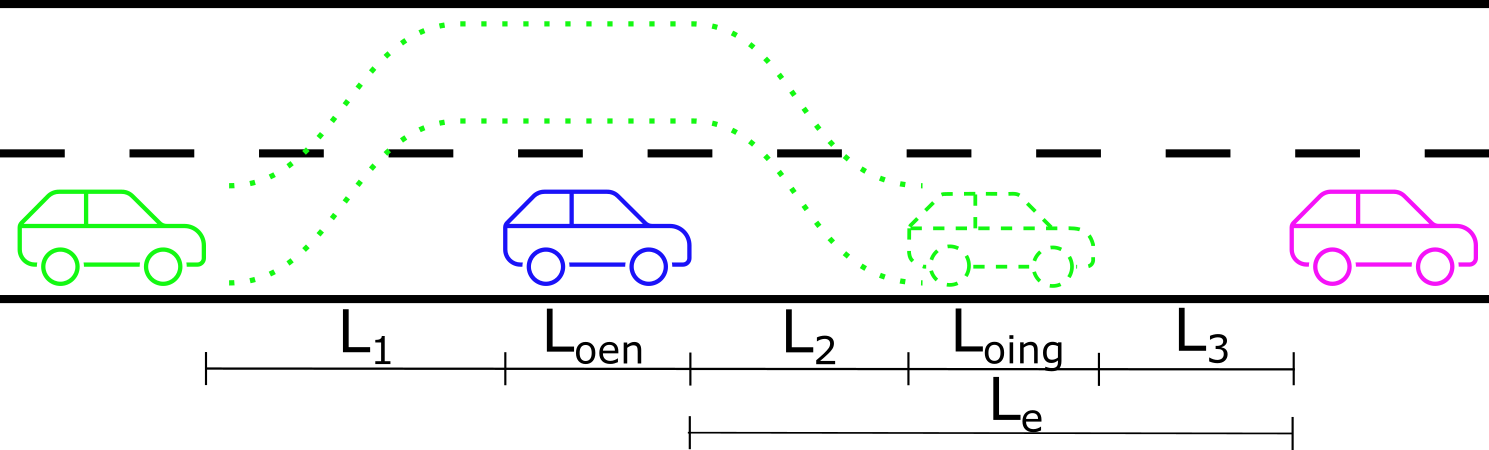}
\caption{Safety distances with another vehicle in front. $L_3$ denotes the distance between the dotted green and the purple vehicle in front. %
The distance between the blue and the purple vehicle is denoted by $L_e$.
}
\Description{This figure shows a schematic of a straight road with cars driving on the right lane where one green car overtakes a blue car and returns to the lane on the right side, between the overtaken blue car and a purple car in front. In addition to the two new distances L_3 and L_e, which are described in the figure caption, various distances known from previous figures are illustrated.}
\label{fig:safety-distance-queue}
\end{figure}

Furthermore, if there is another vehicle in front in phase (iii) of the overtaking maneuver, the overtaking car needs to keep a safety distance to that vehicle, see Figure~\ref{fig:safety-distance-queue}.
We denote this distance by~$L_3$.
Identical to $L_1^s$, we assume $L_3^s=1s$ in our assistant, but this could be user-adjustable. 
If the purple car is beyond the sensor range, it is far enough from the blue car that we can ignore it, as for speeds of up to 120km/h, there will be enough space between the blue and purple car. 
If the purple car is within sensor range (250m) before initiating phase (i), then we treat the blue and purple car as a queue and assume that both drive at the same speed, i.e., %
$v_{oen}$.
In this case, since we set 
$L_2^s = L_3^s=1s$, %
we require
\begin{equation}\label{eq:L_e}
	L_e \geq L_{oing} + 2 \cdot L_2^s \cdot v_{oen}.
\end{equation}

Finally, as it is possible for a driver to abort the overtaking maneuver up to the point-of-no-return (see Section~\ref{subsec:overtaking-beh-model}), we only need to calculate the required sight distance at that point, as suggested in~\cite{NCHRP_NAP23278}.
We predict the point-of-no-return using the acceleration models from Section~\ref{subsec:acc-models} and include the maximum velocity constraint from Section~\ref{subsec:overtaking-beh-model}.
We denote by $t_{rest}$ the time is takes to finish the maneuver from the point-of-no-return, and by $d_{rest}$ the total distance covered during that time. 
The time $t_{rest}$ %
is computed from the acceleration models in Section~\ref{subsec:acc-models}.
This computation also takes into account the maximum speed of the overtaking vehicle and the deceleration as described in Section~\ref{subsec:overtaking-beh-model}.
Moreover, it calculates the speed of the overtaking vehicle at the end of the maneuver, which we denote by $v_{oing,end}$, also considering constraint~\eqref{eq:L_e}.
At the point-of-no-return, only half of the vehicle lengths $L_{oen}$ and $L_{oing}$ need to be considered. 
With all this in mind, the total required sight distance~$d_{s,min}$ is given by:
\begin{equation}\label{eq:reg-sight-dist}
    d_{s,min} = d_{rest} + L_{sm} + d_{opp}
\end{equation}
where
\begin{equation*}
\begin{aligned}
d_{rest} =\ & L_2^s \cdot v_{oen} + \frac{L_{oen}}{2} + \frac{L_{oing}}{2} + v_{oen} \cdot t_{rest} \\
L_{sm} =\ & L_{sm}^s \cdot \left(v_{opp}^{max} + v_{oing,end} \right) \\
d_{opp} =\ & v_{opp}^{max} \cdot t_{rest}
\end{aligned}
\end{equation*}

\subsubsection{Determination of Available Visibility~$d_s$ and Comparison with $d_{s,min}$}\label{subsec:available-sight-distance}
In a last step, the required safety distance~$d_{s,min}$ at the point-of-no-return computed via~\eqref{eq:reg-sight-dist} needs to be compared to the available sight distance~$d_{s}$ at that point.
$d_s$ can be pre-computed using the algorithm described by De Santos-Berbel et al.~\cite{sightDist}, utilizing the Digital Surface Model (DSM). 
We calculate~$d_{s}$ for each point on the road \new{(technically, the road is discretized and values in-between are interpolated)} by determining how far the opposite lane can be seen and store the resulting value for future use. %

\begin{figure}[ht]
\includegraphics[width=\linewidth,trim=0cm 0.55cm 0cm 0cm, clip]{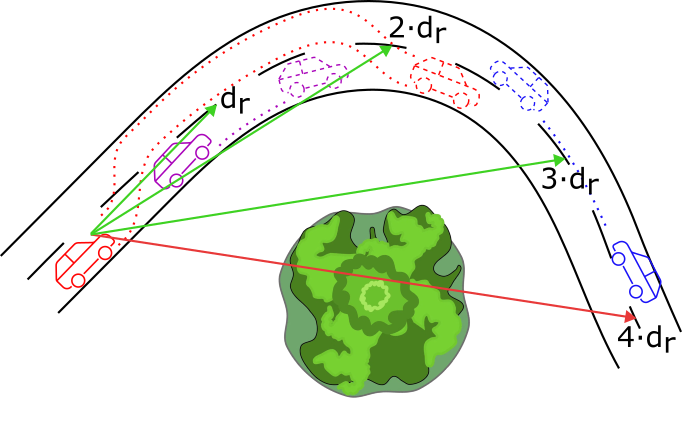}
\caption{\new{Calculating available sight distance for the red overtaking car according to~\cite{sightDist}, taking into account road layout (including slope) and objects that obstruct the view.}\label{fig:avail_sight_distance}}
\Description{This figure shows a schematic of a curved road around a tree. A typical overtaking maneuver is depicted on the road with an overtaking car, an overtaken car and an opposing car. Furthermore, sight lines to the opposing lanes are drawn, some of which the tree blocks.}
\end{figure}

The iterative process of determining how far the opposite lane can be observed from a given position on the road ("eye position") is shown in Figure~\ref{fig:avail_sight_distance}.
\new{
A ray is cast from the eye position (red car) to a point $d_r$ meters ahead on the opposite lane.
If no DSM intersection occurs, another ray is cast, $2\cdot d_r$ meters ahead, then $3\cdot d_r$, until a ray collides with the DSM data (red arrow). 
The last unobstructed distance is then $d_s$.
}
Following NCHRP~\cite{NCHRP_NAP23278}, the eye height and object height are assumed to be 1 m above the road. %

As stated in Section~\ref{subsec:core-concept}, the following cases issue a warning: 
\begin{enumerate}
\item If there is not enough space between two vehicles in front within sensor-range.
\item If $d_s < d_{s,min}$, then the available sight distance is insufficient, hence overtaking is not safe.
\item If $d_s \geq d_{s,min}$ then $d_{s,min}$ is compared to the distance to the nearest intersection~$d_{inter}$ and if $d_{s,min} \geq d_{inter}$, then overtaking is not safe due to an intersection nearby.
\end{enumerate}

\section{Implementation}\label{sec:implementation}

\subsection{Simulator Hardware}\label{subsec:hardware}
\begin{figure}[ht]
\centering
{\includegraphics[height=6.0cm]{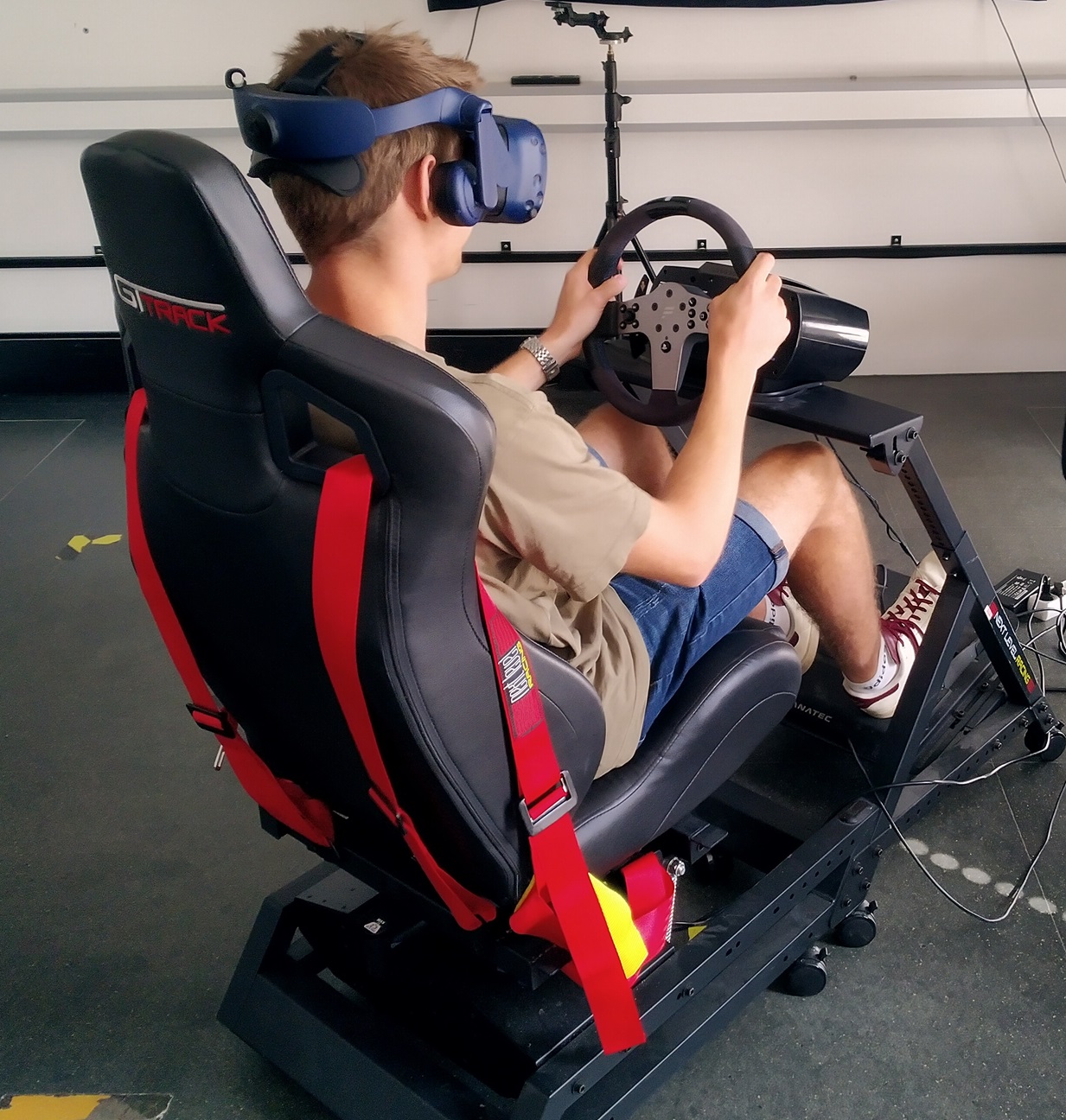}}
\caption{Driving simulator}%
\label{fig:simulator}
\Description{This figure shows a person wearing a VR headset sitting in a race gaming setup featuring a seat with red race seatbelts. The person is using a steering wheel in front and a pedal box down below.}
\end{figure}

The hardware consists of a gaming system (Intel Core i5-9600K, Nvidia RTX 2080Ti), and consistently achieves 100fps frame rate, fast enough for the 90Hz refresh rate of the HTC Vive Pro\footnote{\url{https://www.vive.com/eu/product/vive-pro/}} virtual reality headset.
The driving simulator shown in \autoref{fig:simulator} %
includes a NextLevelRacing GT Track cockpit\footnote{\url{https://nextlevelracing.com}}, equipped with a Fanatec CSL Elite Racing Wheel with brake and gas pedals\footnote{\url{https://fanatec.com}}. 
It also features a movable seat, the NextLevelMotion Platform V3.

\subsection{Virtual Environment}\label{subsec:environment}
The software uses Unity 2021.3\footnote{\url{https://unity.com}} to simulate driving a 140 horsepower front-wheel drive pickup truck with automatic transmission on a country road.
While the car model is from existing assets, we spent considerable time fine-tuning acceleration, center of gravity, and tire forces for more realistic handling.
We include sound effects for braking and airflow at higher speeds. %
While we do not specifically simulate in-car sensors, we derive all information the sensors would yield within their range from the Unity environment.

\begin{figure}
\centering
\includegraphics[width=\linewidth]{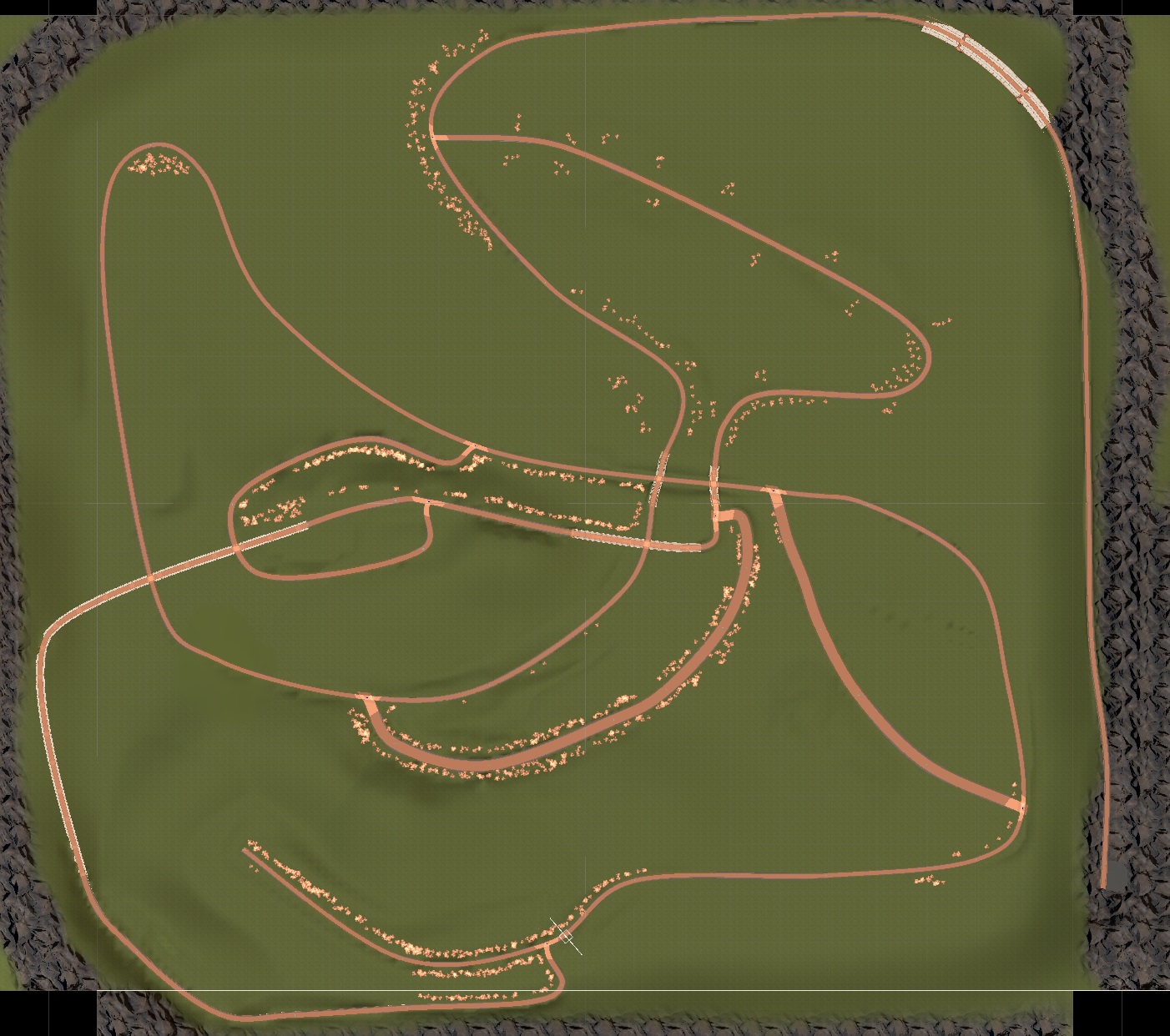}
\caption{Top view of the road network}
\label{fig:road-network}
\Description{This figure shows a road network on green background. Roads and trees surrounding roads are colored orange for better visibility. There are mountains on the edge of the map. The road layout features straight and curvy sections, intersections, and bridges.}
\end{figure}
We manually designed the virtual road network (Figure~\ref{fig:road-network}) using RoadArchitect~\cite{web:roadarchitect}. %
The roads are mostly one lane per direction and have slopes of up to 11.4\%.
The environment includes curves, open fields, hills, and forests that limit visibility. 
For pre-computing available sight distances (Section~\ref{subsec:available-sight-distance}), we approximate the roads with 3D splines parameterized by their length.
\new{Every 10m along the road, we} calculate and store visibility range \new{(with $d_r = 10m$)}, farthest visible point, obstructions, and curvature, treating this as available map data \new{and interpolating values in-between. 
We chose these parameters based on test drives.}

All other vehicles on the roads are autonomous. 
They follow the road network, select random routes at intersections, and adhere to the right-of-way rules.
The target speed of vehicles on the driver's lane are 70-85km/h, depending on the vehicle type and road slope.
Opposing traffic speed can reach up to 100km/h. 

While all three acceleration models from Section~\ref{subsec:acc-models} are implemented, only the \textit{(Constant)} model~\eqref{eq:acc-const} is used for predicting the point-of-no-return and comparing available to required sight distance (Section~\ref{subsec:available-sight-distance}). 
The other models are calculated and logged simultaneously, allowing retrospective evaluation of their prediction accuracy without adding to participants' workload.

\section{User Study}\label{sec:user-study}
To evaluate the utility of the developed assistant and driver acceptance, a user study is conducted in the VR driving simulator. 
Vankov et al.~\cite{VANKOV2021351} provide a survey to assess the potential of VR experiments in road safety. 
While they acknowledge fidelity and validity problems, simulators pose less ethical and safety issues. 
We follow the recommendation of recording a variety of quantitative measurements as listed in Appendix~\ref{app:logdata}.

\subsection{Research Questions and Hypotheses}\label{subsec:hypotheses}
This experiment helps addressing our research questions from Section~\ref{sec:intro}.
As Loewenau et al.'s~\cite{DynamicPassPrediction} passive systems lacks a user study (see Section~\ref{subsec:existing-assis}), we first want to investigate how drivers perceive and accept our passive assistant, and whether and how drivers would actually use it. 
We expect a majority of participants preferring the assistant to driving without and explore this in interview questions and participants' ranking of conditions. 
We also explore the role of personal preferences in open-ended interview questions. 
Cautious about Fank et al.'s~\cite{Frank2021} identified narrow margin between too little and too much information (see Section~\ref{subsec:uis-in-existing-assistants}), we investigate how the two UIs -- one providing minimalistic, the other more detailed information -- differ in SUS and perceived workload, hypothesizing that: 
\begin{enumerate}[label=\textbf{H.1.\arabic*}]
    \item\label{item:rq1-hyp-no-higher-workload} Neither UI has a higher perceived workload than the baseline. %
    \item\label{item:rq1-hyp-monitoring-higher-sus} The monitoring-focused UI will score higher on SUS than the scheduling-focused UI. %
\end{enumerate}

Our second research question examines the safety benefits and potential distractions of our assistant. 
Unlike Walch et al.'s~\cite{Walch2018,Walch2019} autonomous overtaking assistant, we focus on improving drivers' decision-making before overtaking, and promoting a more patient driving style. 
We aim to identify which UI supports this best and analyze the available-to-required sight distance ratio at the point-of-no-return, important for safe overtaking. 
In addition to exploring whether participants feel the assistant enhances overall driving safety, we hypothesize the following:
\begin{enumerate}[label=\textbf{H.2.\arabic*}]
    \item\label{item:rq2-hyp-monitoring-less-distracting} The monitoring-focused UI distracts drivers less than the scheduling-focused UI. %
    \item\label{item:rq2-hyp-follow-longer} With the assistant, drivers follow a vehicle longer compared to the baseline. %
    \item\label{item:rq2-hyp-increase-sight-distance} The assistant increases the median available-to-required sight distance ratio at the measured point-of-no-return compared to the baseline.
\end{enumerate}

Our third research question concerns the accuracy of our sight distance prediction models based on underlying model assumptions. 
We want to investigate if and by how much the more complex acceleration models improve prediction accuracy, hypothesizing that:
\begin{enumerate}[label=\textbf{H.3.\arabic*}]
    \item\label{item:rq3-hyp-ldm-and-dynamic-are-better} Out of the three acceleration models, the "Constant" one will have the most relative error of predicted time to the point-of-no-return with our model assumptions. %
\end{enumerate}

\subsection{Experimental Design and Procedure}\label{study:experimentalDesign}
We consider three conditions, "No Assistant" (Baseline), "Monitoring", and "Scheduling":
\begin{itemize}
	\item "No Assistant" serves as the baseline, representing regular navigation-guided driving, see Figure~\ref{fig:teaser}(A).
	\item "Monitoring" describes navigation-guided driving with the assistant's monitoring-focused UI as described in Section~\ref{subsec:uis}, see Figure~\ref{fig:teaser}(C).
	\item "Scheduling" describes navigation-guided driving with the assistant's scheduling-focused UI as described in Section~\ref{subsec:uis}, see Figure~\ref{fig:ui-variants}.
\end{itemize}

Order effects and carry-over effects are eliminated by Balanced Latin Squares~\cite{Campbell_Geller_1980}. 
Prior to the study, participants were provided with key study information.
After participants read and signed the consent forms and answered demographic questions, they familiarized themselves with the simulator and hardware and performed a test drive without an assistant or traffic. 
If participants felt the motion platform increased motion sickness, it was deactivated. 

Prior to each condition, 
\new{any}
upcoming assistant's functionality was explained and participants could test-drive the current condition for up to five minutes. 
Afterwards, participants drove 28.8 km per condition.
After each condition, a break was issued and participants filled out questionnaires, including the Driving Activity Load Index (DALI)~\cite{dali} and the %
System Usability Scale (SUS). %
Assistant-related questions were omitted in the Baseline condition.
Being cautious about motion sickness, in each condition we split the 28.8 km journey into two routes (16.1 km and 12.7 km) with a random order to allow for a break also within a condition.

\new{
Participants were not required to overtake and the experiment did not rely on participants' overtaking preferences. 
They were instructed to drive as they would in reality, which we reiterated before each condition. 
We did not orchestrate overtaking opportunities, with one exception, which prevented following a convoy for too long: 
Since our assistant assumes overtaking one car at a time (Section~\ref{subsec:overtaking-beh-model}), %
if a convoy of vehicles formed and the driver followed it for 30 seconds, all AI vehicles except the one ahead would signal turning right and "park" on the roadside.
Depending on the available sight distance and oncoming traffic, this gave the driver the opportunity to overtake the remaining vehicle.}

During the experiment, we logged the driving data into a CSV file -- for a detailed list, we refer to Appendix~\ref{app:logdata}. 
In addition, we recorded the participant's view using OBS\footnote{\url{https://obsproject.com}}.
After participants completed all three conditions, they were asked about their experience in a semi-structured interview. 
This included ranking the conditions, %
if they felt distracted by the assistant, if the assistant made them more patient or pressured to overtake, if the assistant improved overtaking safety, in which situations the assistant would be helpful, and if they have any suggestions for improvements.

The study concluded with a debriefing and 25 Euro compensation. 
The entire process, including instruction, practice drives, main drives, questionnaires, and interviews, usually took about 2 hours, occasionally extending up to 3. %

\subsection{Participants}
25 individuals participated in the study, with 20 completing it (15 males, 4 females, 1 diverse, ages 19-60, average age 27.4). 
All 25 received financial compensation. 
Of the 20 who completed it, all except P9 hold a driver's license, with one having a truck license and four having a motorcycle license in addition.
Since the perspective of a single, unbiased, inexperienced driver is anecdotal, we disregard P9's data entirely.
On average, the remaining 19 participants have held their licenses for 10.6 years (std: 8.8; min: 2; max: 44) and have driven 9105 km (std: 9301; min: 0; max: 35000) in the past 12 months. 
All participants had previous experience with 3D video games and 14 had previous experience with driving simulators. %
\new{The majority used the motion platform throughout.}
For other questions participants answered before driving we refer to Appendix~\ref{app:participants}.

\section{Open Science}\label{sec:open-science}
During the experiment, we logged extensive driving data from each participant who completed the study -- 1240 minutes of driving %
and 576 completed overtaking maneuvers -- %
into CSV files (6.5 GB in total). 
These constitute TOWARDS, The Overtaking Warning Assistant Recorded Data Set~\cite{dataset:towards}, publicly available at \url{https://doi.org/10.5281/zenodo.14757143}. 
We detail its contents in Appendix~\ref{app:logdata}.
Moreover, we release the source code for the safe sight prediction models from Section~\ref{subsec:prediction-model}, for easy inclusion in Unity environments, available at \url{https://github.com/abauske/YouShallNotPass}.

\section{Results}\label{sec:results} %
We report and analyze the results from our user study, highlighting all statistically significant findings. 
Statistical tests were conducted using one-factor ANOVA, with t-tests for normally distributed data (validated via the Shapiro-Wilk-Test~\cite{shapiro}). 
For non-normal data, we used the Wilcoxon signed-rank test~\cite{wilcoxon} and the Mann-Whitney U Test~\cite{mannwhitneyU}. 
Effect sizes are given for significant results. 
We use Glass's $\Delta$ if the standard deviations deviate by more than 20\%. 
Otherwise, we use Cohen's $d$ for equal sample sizes over 20, and Hedge's $g$ for smaller sample sizes.

\begin{figure*}[!th]
\includegraphics[width=0.85\linewidth]{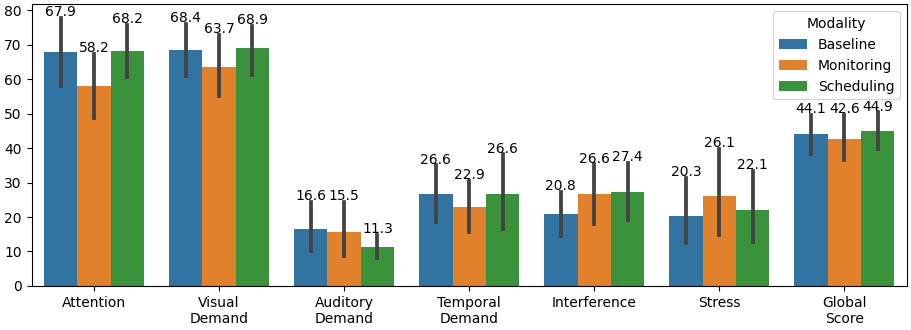}
\caption{Participants' reported workload in the DALI questionnaire (mean and 95\% confidence interval). The global score is computed with equally weighted categories. 
\Description{This figure shows a bar plot with the following categories on the x axis: "Attention", "Visual Demand", "Auditory Demand", "Temporal Demand", "Interference", "Stress", and "Global Score". For each of these categories there is a bar for each of the conditions "Baseline", "Monitoring", and "Scheduling". Attention and Visual Demand ratings are highest at around 60 with the Monitoring condition always being lowest, and Baseline and Scheduling being equally high. The Global Score is similar between conditions (Baseline: 44.1, Monitoring: 42.6, Scheduling: 44.9), with similar standard deviation. Temporal Demand, Interference and Stress have values between 20 and 30 for all conditions. For Interference, Baseline is lowest and the others are on the same height. For Stress, Baseline is lowest and Monitoring is highest. Lowest values are given for Auditory Demand, ranging from 11 to 17, with Baseline highest and Scheduling lowest.}
\label{fig:dali}}
\end{figure*}
\subsection{Accidents Happen -- Regardless of Assistants} %
In our study, all except one collision were unrelated to overtaking.
Poor AI performance at intersections and random placement of moving vehicles at the start of the drive led to 9 collisions. 
Drivers misjudging turns or failing to maintain lanes caused 4 collisions. 
In the remaining accident, P6 ignored the assistant's warning (mon\-i\-tor\-ing-focused UI) and attempted to overtake two cars at once. 
They overlooked oncoming traffic, with which they collided sideways.

\subsection{\new{Participants rate the Baseline (No Assistant) worst but are divided on the (second-)best UI}}\label{subsec:assis-yay-or-nay}
\begin{figure}[ht]
\includegraphics[width=\linewidth]{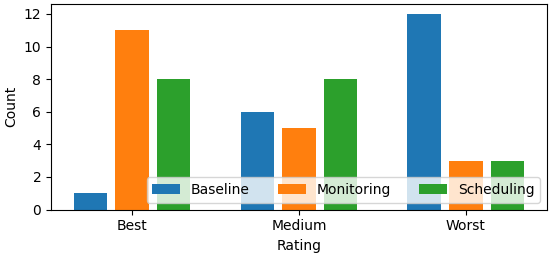}
\caption{Participants' rating of the conditions. Rating multiple conditions equally was permitted. \label{fig:assistant_rating}}
\Description{This figure shows a bar plot with "Count" on the y axis, and x axis consisting of three "Rating" options: "Best", "Medium", and "Worst". For each of these three options there are three bars: Blue for "Baseline" (Best: 1; Medium: 6; Worst: 12), orange for "Monitoring" (Best: 11; Medium: 5; Worst: 3) and green for "Scheduling" (Best: 8; Medium: 8; Worst: 3).}
\end{figure}
At the end of the study, participants were asked to rank the three conditions "No assistant" (Baseline), "Monitoring", and "Scheduling" from "Best" to "Worst". 
We allowed equal rating of conditions. 
The responses are illustrated in Figure~\ref{fig:assistant_rating}.
While participants were divided on which UI is (second-)best, the Baseline condition was rated "worst" the most %
and rated "best" only once by P10, who commented their choice with \textit{"being familiar with driving without an assistant"}. 
While driving in the Baseline condition after having experienced the assistant in the "Scheduling" condition, %
P16 stated, \textit{"You only know what you've got when you've lost it. Before, I had the impression I didn't really pay attention to it, but now I realize how I would like it to be available."} %

\subsection{Driver's Workload: The monitoring-focused UI distracts drivers less and requires the least attention}\label{subsec:dali}
We evaluated driver's workload with the Driving Activity Load Index (DALI), which stemmed from NASA-TLX but was adapted to the driving task in~\cite{dali}. 
Figure~\ref{fig:dali} shows the participants' subjective assessment.

While the equally weighted global score shows no significant difference between conditions, %
"Attention" and "Visual Demand" stand out with the highest workload values (58 to 69). 
The monitoring-focused UI requires significantly less attention than the scheduling-focused UI ($p=0.04$, Wilcoxon, medium effect Hedges's $g=0.50$) and less than the baseline ($p=0.09$, Wilcoxon, small effect Hedges's $g=0.44$). 
Visual demand is also lower with the monitoring-focused UI compared to the baseline ($p=0.06$, Wilcoxon, small effect Hedges's $g=0.26$). 
Other workload categories have lower values (11 to 27) without significant differences. 
The auditory demand is lowest in all conditions. 
We recall that, besides braking and airflow sound, there was no audio. %
In total, we see no support for~\ref{item:rq1-hyp-no-higher-workload}.

\begin{figure}[htb]
\includegraphics[width=\linewidth]{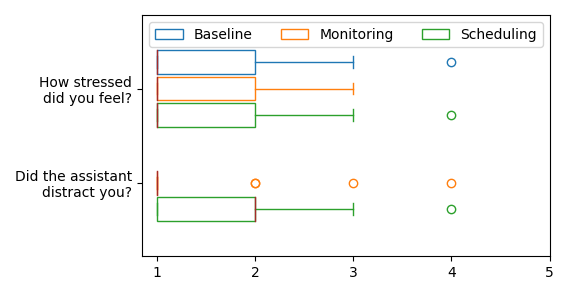}
\caption{Boxplot of the 5-point Likert scale analysis of questions related to workload. %
\label{fig:workload}}
\Description{The figure presents a boxplot analysis of responses to workload-related questions using a 5-point Likert scale (1 to 5). The plot compares three conditions: Baseline, Monitoring, and Scheduling. The first question is, "How stressed did you feel?". For "Baseline", "Monitoring" and "Scheduling", the values are mostly equal, ranging from 1 to 3 with the box from 1 to 2 and a median of 1. The second question is, "Did the assistant distract you?". This question omits the "Baseline" condition. For "Monitoring", all values except three outliers at 2, 3 and 4, are "1". For "Scheduling", values range from 1 to 3 with the box from 1 to 2 and a median of 2.}
\end{figure}
\begin{figure*}[ht]
\includegraphics[width=0.8\linewidth]{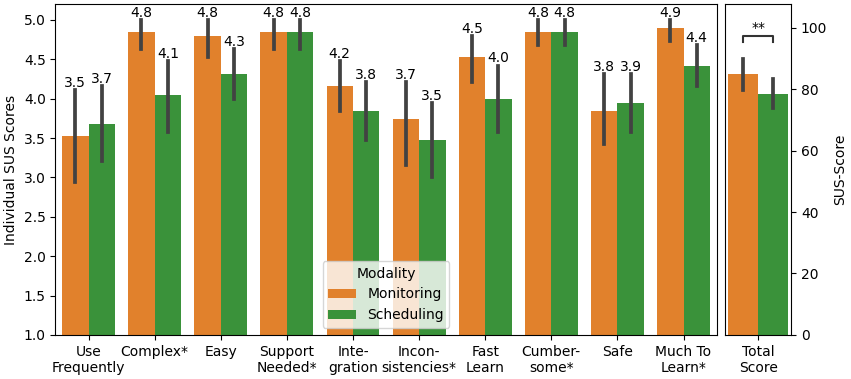}
\caption{System Usability Scale: Individual categories and total score for both UI variants (mean and 95\% confidence interval). Individual categories marked with $^*$ have inverted scores, i.e., higher scores are always better. 
\label{fig:sus}}
\Description{This barplot shows the mean and 95\% confidence interval of the SUS individual scores and the SUS total score for the two conditions "Monitoring" and "Scheduling". Higher scores are always better. The average values for the individual categories are as follows ("Monitoring" vs. "Scheduling"): 3.5 vs. 3.7; 4.8 vs. 4.1; 4.8 vs. 4.3; 4.8 vs. 4.8; 4.2 vs 3.8; 3.7 vs. 3.5; 4.5 vs. 4.0; 4.8 vs. 4.8; 3.8 vs. 3.9; 4.9 vs. 4.4. The total score is significantly (**) higher for the "Monitoring" condition.}
\end{figure*}

We asked participants about whether and how the assistant distracted them, both on a Likert scale after each condition (see Figure~\ref{fig:workload}) and as an open question in the interview. 
Participants found the monitoring-focused UI significantly less distracting than the scheduling-focused UI ($p=0.008$, Wilcoxon, medium effect Hedges's $g=0.69$), supporting~\ref{item:rq2-hyp-monitoring-less-distracting}.
They reported the changing distance numbers (updated every second) as a major distraction, deeming them inconsistent.
In interviews, participants stated that the assistant, irrespective of UI, takes many thoughts off their minds (%
\textit{"You have to look less with the assistant because otherwise you're always looking for an overtaking opportunity"} (P4)) and gives them more time to prepare for overtaking (\textit{"With the numbers [scheduling-focused UI], you could prepare better and not be surprised by an overtaking opportunity"} (P7)).

\subsection{System Usability\new{: Most participants would almost never turn off either UI, but the monitoring-focused UI scores higher on SUS}}\label{subsec:sus}
Figure~\ref{fig:sus} shows the mean and 95\% confidence interval for both the individual categories of the System Usability \new{Scale} and the total score of both UIs.
The monitoring-focused UI receives a significantly higher total score (85.0 vs 78.6; $p=0.006$, %
paired t-test, medium effect Hedges's $g=0.57$).
Both scores exceed the typical 68-70 points reported in the literature~\cite{sus_retrospective,interpreting_sus}, placing them in the acceptable range with grades between A~\cite{interpreting_sus} and B~\cite{sus_retrospective}, depending on the scale used.
In the individual categories, the monitoring-focused UI scores significantly better than the scheduling-focused UI in "complexity" (%
$p=0.008$, Wilcoxon, very large effect Glass's $\Delta=1.57$), "easy-to-use" (%
$p=0.03$, Wilcoxon, large effect Glass's $\Delta=0.89$), "fast-to-learn" (%
$p=0.008$, Wilcoxon, medium effect Glass's $\Delta=0.76$), and "much-to-learn" (%
$p=0.007$, Wilcoxon, very large effect Glass's $\Delta=1.50$), with no further significant differences.
This supports~\ref{item:rq1-hyp-monitoring-higher-sus}.

\begin{figure}[ht]
\includegraphics[width=\linewidth]{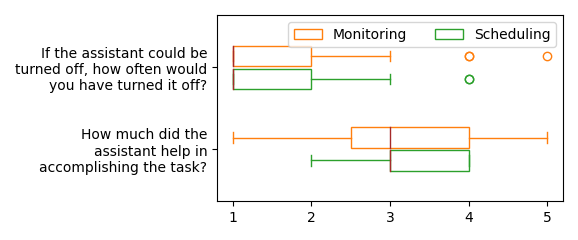}
\caption{Boxplot of the 5-point Likert scale analysis of questions related to the perceived usefulness of the assistant. The task description to perform a navigation-guided drive on a country road as in reality was repeated before the question.} \label{fig:usability}
\Description{The figure presents a boxplot analysis of responses to questions related to perceived usefulness of the assistant using a 5-point Likert scale (1 to 5). For the first question ("If the assistant could be turned off, how often would you have turned it off?"), the two whiskers (almost) coincide, with the "Monitoring" condition having an additional outlier at "5". The range of responses for the second question ("How much did the assistant help in accomplishing the task?") spans the whole Likert scale and is much larger for the Monitoring condition.}
\end{figure}

Other questions about the assistant's perceived usefulness (Figure~\ref{fig:usability}) showed no significant differences. 
Notably, most participants reported they would almost never turn off either variant. %

\subsection{Individual UI preferences \new{play a major role on perceived usefulness}}\label{subsec:indiv-ui-prefs}

Comments during driving and post-study interviews highlight participants were divided on the UIs. 
Some favored the scheduling-focused UI for feeling more patient and relaxed (P1, P2, P6, P7, P8), its additional warning icon for too little distance between two vehicles ahead (P13, P21), and the distance to the next lane opening (P13). 
Others preferred the monitoring-focused UI (P3, P4, P12, P14, P16, P17, P19, P20, P23) for reasons like lower mental load, minimal distraction (P3, P12, P16), and feeling more relaxed and patient (P4, P20). 
A car dealer participant noted older drivers (who are more likely to buy more modern and expensive cars) would appreciate the simplicity of the monitoring-focused UI, as they often struggle with the increasing number of driving assistants and their many (blinking) icons.

We also found a few cases in which the non-preferred UI is detrimental to driving experience and performance.
P11 noted that the scheduling-focused UI made them consider whether overtaking was feasible despite the warning, comparing it to a "game" of beating a navigation device's estimated arrival time. 
P3 made a comparable comment.
Similarly, P13 stated the monitoring-focused UI made them question the system, tempting them to \textit{"compete against the system and overtake despite the warning"} since it lacked explanation.

While participants reported \new{in their interviews and questionnaires} the overtaking maneuver itself is not made safer by the assistant (for more questions on perceived safety, see Appendix~\ref{app:perceived-safety}), participants felt the assistant (in their preferred UI) does make the overall driving safer by facilitating the decision of whether to overtake or not (P8, P12, P14, P17, P19, P21, P23) and by enabling a more patient driving style (P1, P2, P3, P4, P6, P7, P8, %
P12, P16, P19, P20).

When asked in what situations they could imagine using the assistant in their preferred UI, some participants would mainly use it on unfamiliar roads (P1, P2, P5, P6, P10, P13, P14, P20, P23), i.e., situational, while others would use it on all country roads (P4, P7, P8, %
P11, P12, P16, P17, P19, P21), i.e., permanently.
Cross-referencing this with the preferred UI yields that a majority of those who prefer the scheduling-focused UI would use the assistant situationally, while the majority of those who prefer the monitoring-focused UI would use it permanently.

\subsection{Driving Data\new{: While the assistant does not increase the available-to-required sight distance ratio, drivers follow vehicles longer}}\label{subsec:driving-data}

\begin{figure}[ht]
\includegraphics[width=0.9\linewidth]{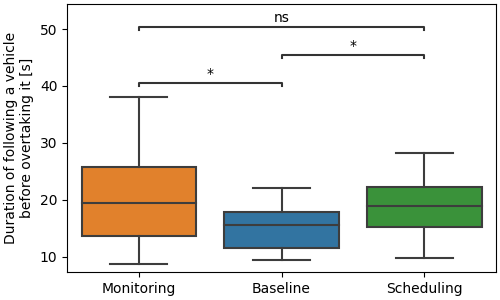}
\caption{Boxplot of the duration a driver followed a vehicle for the three conditions.}
\label{fig:following-duration}
\Description{The figure is a boxplot depicting the duration a driver followed a vehicle before overtaking under three conditions: Monitoring (orange), Baseline (blue), and Scheduling (green). The statistical significance of differences between conditions is indicated above the plot. 
Axes: The y-axis is labeled "Duration of following a vehicle before overtaking it [s]" and ranges from 10 to 50 seconds, with increments of 10 seconds. The x-axis shows the three conditions: Monitoring, Baseline, and Scheduling.
Boxplot Details: Each condition is represented by a vertical boxplot. The boxes for the three conditions are color-coded: orange (Monitoring), blue (Baseline), and green (Scheduling). The horizontal line inside the box represents the median time.
Observations: The "Monitoring" condition exhibits the widest range (10 to 40) and highest median following duration. The "Baseline" shows a shorter median following duration compared to Monitoring. The "Scheduling" condition has a median almost identical to "Monitoring", with a smaller range (10 to 30). A significant difference (*) exists between Monitoring and Baseline. Another significant difference (*) is observed between Baseline and Scheduling. The difference between Monitoring and Scheduling is not significant (ns).}
\end{figure}

The driving data in Figure~\ref{fig:following-duration} shows participants followed a vehicle significantly longer with either UI (Mann-Whitney U; Monitoring: $p=0.03$, Scheduling: $p=0.01$) at medium effect sizes (Monitoring: Hedges's $g=0.59$, Scheduling: Glass's $\Delta=0.75$), with no significant difference between the UIs.
We thus see support for~\ref{item:rq2-hyp-follow-longer}.
\new{Meanwhile, the average number of overtakes did not vary greatly between conditions (Baseline: 9.7, Monitoring: 10.4, Scheduling: 9.5).}

\begin{figure}[ht]
\includegraphics[width=0.9\linewidth]{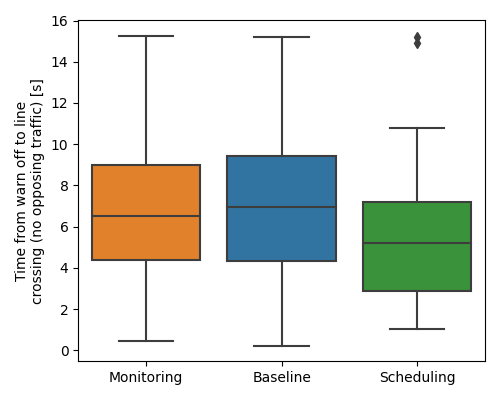}
\caption{\new{Time between warning turning off and crossing the middle line for situations where participants overtook after the warning disappeared and there was no oncoming traffic.}
\label{fig:time-between-warning-off-and-overtaking}}
\Description{The figure is a boxplot depicting the time from the warning turning off to participants crossing the middle line (measured in seconds) in overtaking situations where participants overtook after the warning disappeared and there was no oncoming traffic. It compares three experimental conditions: Monitoring, Baseline, and Scheduling.
Axes: The y-axis is labeled "Time from warn off to line crossing (no opposing traffic) [s]" and ranges from 0 to 16 seconds, with increments of 2 seconds. The x-axis shows the three conditions: Monitoring, Baseline, and Scheduling.
Observations: The medians for "Monitoring" and "Baseline" are similar, around 6–8 seconds, while "Scheduling" has a slightly lower median, closer to 5 seconds. The interquartile range for "Scheduling" is narrower, indicating less variability compared to the other two conditions. There are notable outliers in the "Scheduling" condition, which reach approximately 15 seconds. }
\end{figure}

\new{
To investigate whether the warning disappearing triggered an immediate initiation of the overtaking maneuver, 
we consider situations where participants overtook after the warning disappeared and there was no opposing traffic. 
Note that even though participants never saw a warning in the Baseline condition, we did log whether a warning would have been shown. 
In Figure~\ref{fig:time-between-warning-off-and-overtaking} we notice a trend towards less variance for the scheduling-focused UI and, more importantly, we see that participants did not tend to overtake directly once the warning was gone.
}

\begin{figure}[ht]
\includegraphics[width=0.95\linewidth]{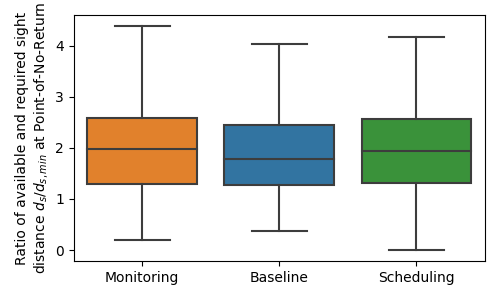}
\caption{Boxplot of available divided by required sight distance at the measured point-of-no-return for the three conditions.}
\label{fig:sight-distance}
\Description{The figure is a boxplot depicting the the ratio of available sight distance to required sight distance (d_s / d_{s,min}) at the point-of-no-return under three conditions: Monitoring (orange), Baseline (blue), and Scheduling (green).
Axes: The y-axis is labeled "Ratio of available and required sight distance d_s / d_{s,min} at point-of-no-return" and ranges from 0 to 4.5 seconds, with increments of 1. The x-axis shows the three conditions: Monitoring, Baseline, and Scheduling.
Observations: Across all conditions, the median is around 2. There are no substantial differences between the three conditions, as their distributions appear quite similar (all ranging from almost 0 to 4.5 with box between 1.2 and 2.6).}
\end{figure}

The sight distance at the point-of-no-return is crucial for evaluating overtaking safety. 
Figure~\ref{fig:sight-distance} shows the ratio of available to required sight distance at the \textit{measured} point-of-no-return\new{, i.e., when the participant's car's center aligned with that of the vehicle to be overtaken}. 
While the sight distance is mostly sufficient across all conditions, some maneuvers with insufficient sight distance (ratios below 1) occurred in each condition. 
No significant differences were found between conditions, offering no support for~\ref{item:rq2-hyp-increase-sight-distance}. 
After thoroughly examining the driving data, we discovered the assistant did not always predict the point-of-no-return accurately, \new{i.e., it did not always match the measured point-of-no-return. This} further motivated the subsequent model comparison to driving data. 

\begin{figure*}[ht]
	\includegraphics[width=0.8\linewidth]{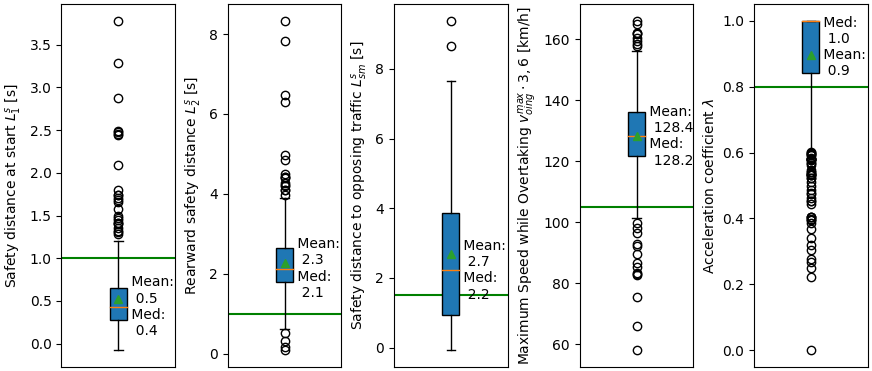}
	\caption{Comparison of model parameters $L_1^s$, $L_2^s$, $L_{sm}^s$, $v_{oing}^{max}$, and $\lambda$ to measured driving data}
	\label{fig:model-assumptions}
    \Description{The figure consists of five boxplots of measured data regarding the 5 model parameters mentioned in the caption. Green lines show the assumed model parameters, outliers are marked as black dots. This plot is described in more detail in the paper text.}
\end{figure*}
\new{Looking at the overtaking maneuvers themselves (Appendix~\ref{app:qual-overtaking}), we found no qualitative differences 
between conditions regarding 
(i) the maximum speed during overtaking (Figure~\ref{fig:safety-app-1}), 
(ii) the duration on the overtaking lane (Figure~\ref{fig:safety-app-2}), and
(iii) the distance to the overtaken vehicle after reeving (Figure~\ref{fig:safety-app-3}).
While Figure~\ref{fig:safety-app-4} might indicate a trend toward more safety distance to opposing traffic with the monitoring-focused UI, this data could have been affected by the prediction error of the-point-of-no-return.}

\subsection{Model Comparison and Key Model Assumptions\new{: If we anticipate drivers speeding during overtaking, we get near-zero relative error in predicting the point-of-no-return}}\label{subsec:model-comparison}
In order to provide accurate warnings before drivers initiate an overtaking maneuver, it is important that models reflect actual observations. 
In particular, %
the point-of-no-return %
should be predicted correctly. %
Otherwise, our assistant will compare the wrong sight distances.
The prediction models require several parameters. 
The following five parameters are present in all models and were chosen carefully based on literature, see Section~\ref{subsec:prediction-model}.
The comparison to measured data is illustrated in Figure~\ref{fig:model-assumptions}. 
The green horizontal line represents the model constants, while the measured values are shown in the respective box plot.

The first two plots show safety distances (in seconds) to the vehicle being overtaken. 
While swerving safety distance ($L_1^s$) is considerably lower than the assumed average (0.5s vs 1s), the greatest visibility is needed at the point-of-no-return, by which time $L_1$ is already covered. 
The reeving safety distance ($L_2^s$) is about twice as high, meaning drivers established  more distance to the overtaken vehicle before reeving than anticipated.
The crucial safety distance to an oncoming vehicle ($L_{sm}^s$) is greater than assumed (1.5s) in most, though not all cases, making it a good fit.
As the chosen assumption is already at the lower end of the values given in literature (Section~\ref{subsec:calculating-safety-distances}), it should not be reduced further.

\begin{figure}[ht]
\includegraphics[width=\linewidth]{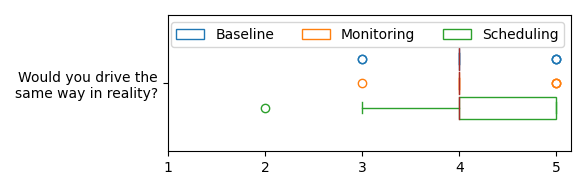}
\caption{Boxplot of the 5-point Likert scale analysis of the question "Would you drive the same way in reality?".} %
\label{fig:Q8}
\Description{This boxplot compares the three conditions regarding the question mentioned in the caption. Regardless of condition, "4" is the most chosen option. Baseline and Monitoring conditions have much less variance.}
\end{figure}

The key finding is in the last two plots of Figure~\ref{fig:model-assumptions}.
Participants overtook at much higher maximum speeds (averaging 128 km/h) and often kicked-down the gas pedal ($\lambda=1$).
During driving, participants stated they want to finish overtaking as fast as possible and kick down the gas pedal also in reality, and confirmed (see Figure~\ref{fig:Q8}) that they followed our instruction to drive in the simulator as they would in reality.
Comments during driving such as \textit{"I have just tried to put my elbow down. For the third time."} (P16), \textit{"I always reach into the air because I want to activate the turn signal lever."} (P19), and similar comments from other participants %
reinforce this.

\begin{figure}[ht]
\includegraphics[width=\linewidth]{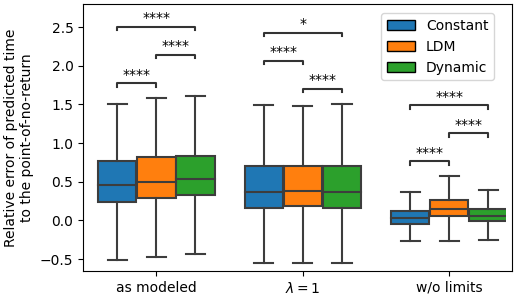}
\caption{Comparison of relative error of duration up to the point-of-no-return. Modeled results ("as modeled") compared to postcomputed values with full acceleration ("$\lambda = 1$") and full acceleration without maximum speed limit ("w/o limits"). 
Values $> 0$ mean the predicted point-of-no-return is reached later than the measured one. Arranged by model assumptions.} %
\label{fig:eq_time-comparison-transposed}
\Description{This boxplot displays the relative error of the predicted time to the point-of-no-return. On the x-axis, the acceleration models are grouped by model assumptions ("as modeled", "lambda=1", and "w/o limits"). For "as modeled" and "lambda=1", the visual differences are minor, albeit significant differences, with ranges between -0.5 to 1.6 with box between 0.2 and 0.8). For "w/o limits", the ranges and boxes reduce for all models (range from -0.3 to 0.6, with box from -0.1 to 0.3), and the LDM model has the largest error. }
\end{figure}

Both the maximum velocity and the acceleration are key in predicting the point-of-no-return.
Hence, we examine the relative error in predicted time %
to the measured point-of-no-return with the original and adjusted assumptions for the three acceleration models.
For our model assumptions ("as modeled"), Figure~\ref{fig:eq_time-comparison-transposed} shows significant differences between the acceleration models (Wilcoxon; $p<0.001$ throughout), albeit with very small effect sizes (Cohen’s $d$ between $0.028$ and $0.061$).
Yet, since the "Constant" model has less relative error (it performs well even with adjusted assumptions), hypothesis~\ref{item:rq3-hyp-ldm-and-dynamic-are-better} is not supported. %
For a complete list of effect sizes and $p$-values, we refer to Appendix~\ref{app:model-comp-effect-sizes}.

\begin{figure}[ht]
\includegraphics[width=\linewidth]{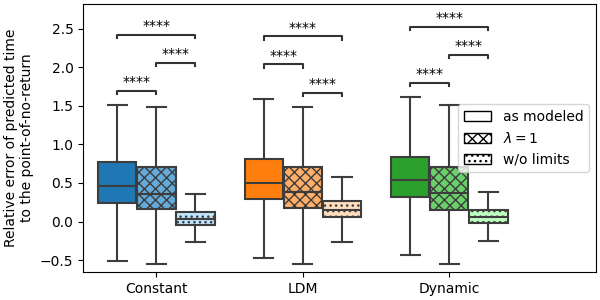}
\caption{
Same data as in Figure~\ref{fig:eq_time-comparison-transposed} but arranged by acceleration model.}
\label{fig:eq_time-comparison}
\Description{This boxplot displays the same data as Figure 21, but grouped by acceleration model, focusing on comparing the model assumptions within a model. For all three acceleration models, the relative error reduces slightly for lambda=1 and approaches zero for "w/o limits".}
\end{figure}

Arranging the same data by acceleration model, Figure~\ref{fig:eq_time-comparison} shows that increasing $\lambda$ to one significantly reduces relative error across all models (Wilcoxon; $p<0.001$ throughout), though with (very) small effect sizes (Constant: Glass's $\Delta=0.156$; LDM: Cohen’s $d = 0.157$; Dynamic: Cohen’s $d = 0.211$). 
Additionally lifting the maximum speed limit brings all acceleration models to near-zero relative error and significantly better performance compared to "as modeled" (Wilcoxon; $p<0.001$ throughout), with huge to very large effect sizes ($1.628 < Glass's~\Delta < 2.405$).
While the results demonstrate the usefulness of even the simplest acceleration model, lifting the maximum speed limit remains questionable. 
Should the assistant really plan for a traffic violation?

\subsection{\new{Robustness Evaluation of Sight Distance Computation}}\label{subsec:robustness}
\begin{figure*}[ht]
\centering
\subcaptionbox{\new{Ego-car placement every 10m; $d_r = 10m$ (Baseline)}\label{fig:400m-baseline}}%
[.45\textwidth]{\includegraphics[width=\linewidth]{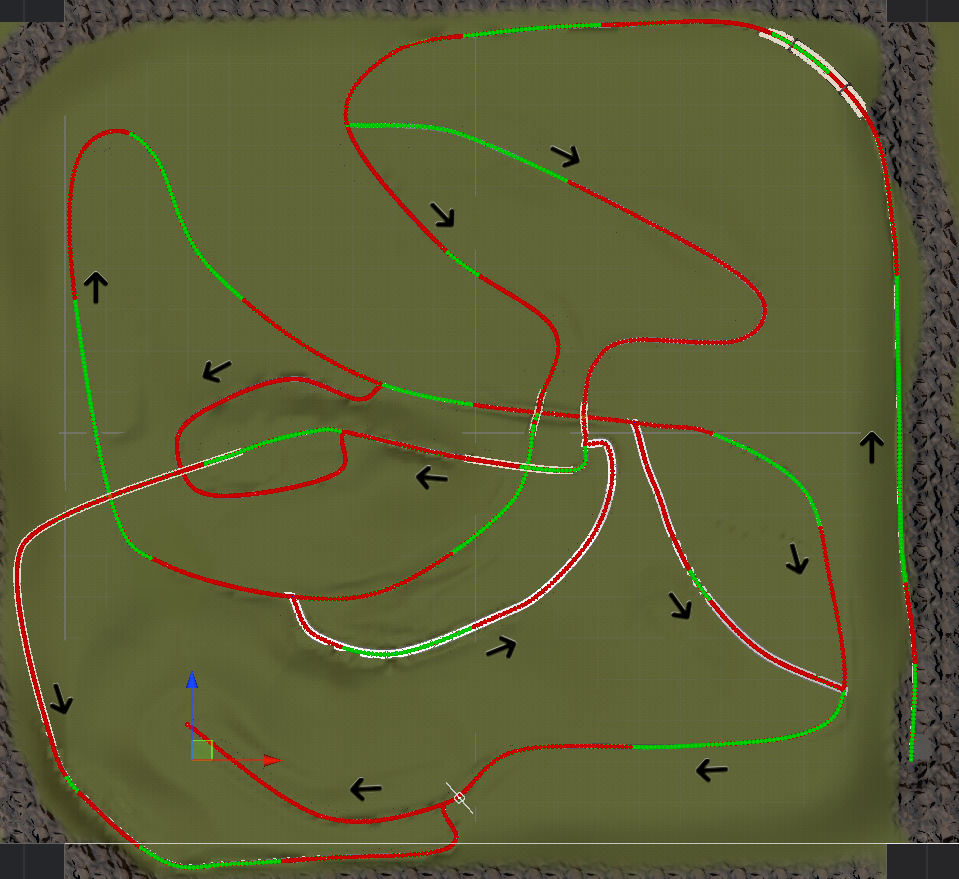}}
\subcaptionbox{\new{Ego-car placement every 1m; $d_r = 1m$}\label{fig:400m-1x1}}%
[.45\textwidth]{\includegraphics[width=\linewidth]{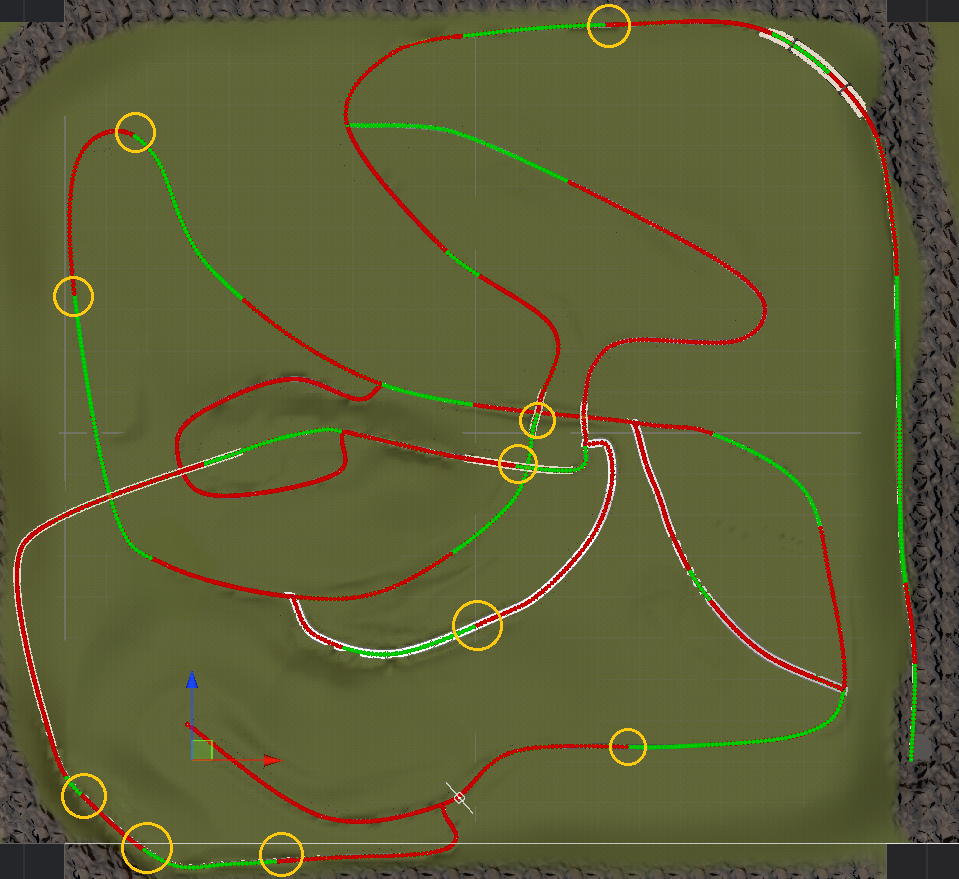}}
\subcaptionbox{\new{Ego-car placement: every 10m; $d_r = 60m$}\label{fig:400m-10x60}}%
[.45\textwidth]{\includegraphics[width=\linewidth]{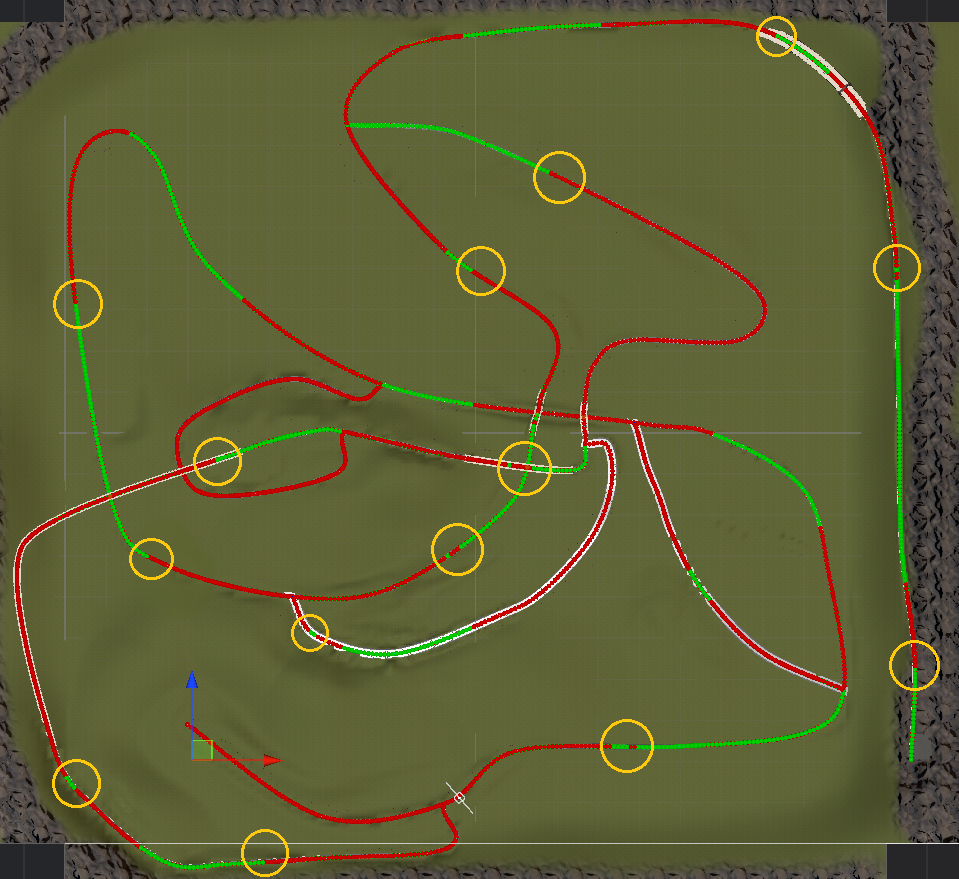}}
\subcaptionbox{\new{Ego-car placement: every 30m; $d_r = 60m$}\label{fig:400m-30x60}}%
[.45\textwidth]{\includegraphics[width=\linewidth]{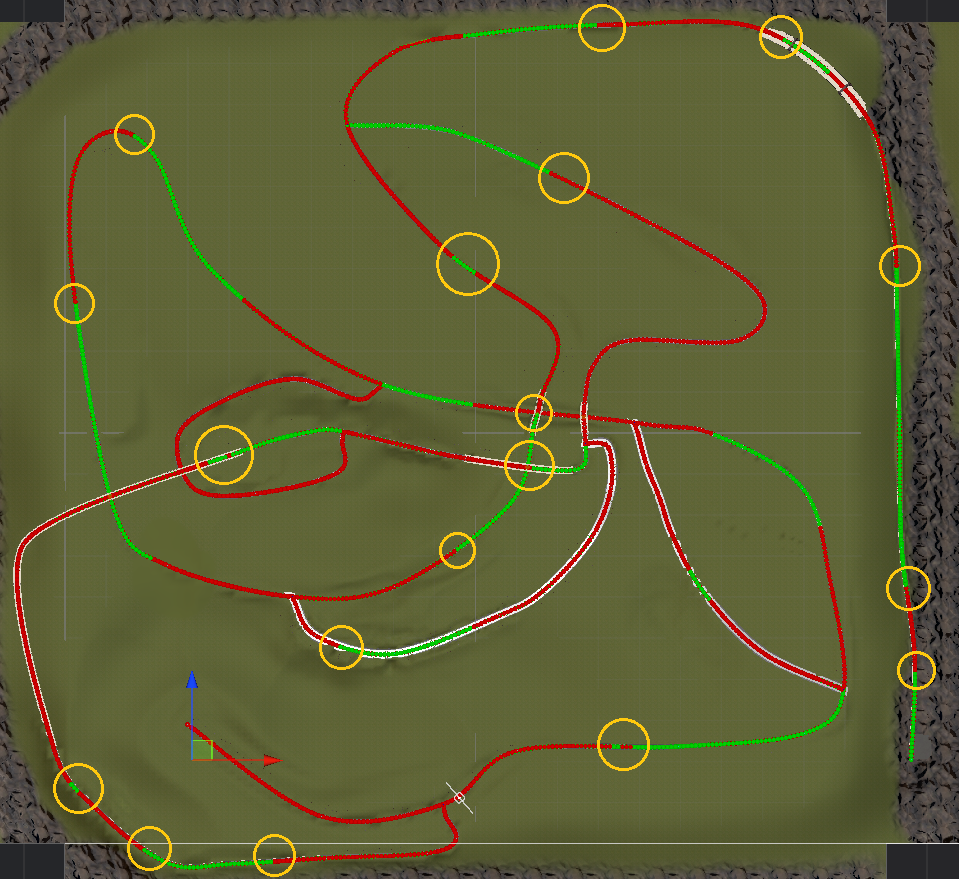}}
\caption{\new{Road layout with indication of 400m available sight distance (green$=$yes, red$=$no) along the driving direction (black arrows) for various intervals of ego-car placement and raycasting ($d_r$). Yellow circles mark changes compared to the baseline (top left).}}
\label{fig:400m-sight-distance-raycast-egocar-params}
\Description{This figure consists of 4 subfigures. All of them illustrate a top-down map showing the same road layout of the virtual road network, with indications of 400m available sight distance (visibility) along the driving direction. The road network appears curvy and includes various turns, loops, and straight sections. Green segments indicate road sections where the available sight distance is at least 400 meters. Red segments represent areas where the sight distance is less than 400 meters. The subfigures visualize how sight distance changes depending on the ego-car's placement and a raycasting interval.
(a) This is the baseline figure. In this configuration, the ego-car is placed at intervals of 10 meters along the road. Raycasting is also performed at 10-meter intervals. Black arrows are overlaid on the map to illustrate the driving direction along the road network. Green segments dominate the straighter sections of the road, where long sight distances are generally available. Red segments occur predominantly in sharp curves or areas with potential obstructions.
(b) In this configuration, the ego-car is placed at intervals of 1 meters along the road. Raycasting is also performed at 1-meter intervals. Visually, it looks similar to the baseline. 10 yellow circles are overlaid on the map to mark changes compared to the baseline configuration (a). All of them are located where the road transitions from adequate (green) to limited (red) visibility or vise versa.
(c) In this configuration, the ego-car is placed at intervals of 10 meters along the road. Raycasting is performed at 60-meter intervals. Visually, it looks similar to the baseline. 14 yellow circles are overlaid on the map to mark changes compared to the baseline configuration (a). Most of them are located where the road transitions from adequate (green) to limited (red) visibility or vise versa, usually prolonging the red segments. Sometimes, a short red segment appears within a green segment and vice versa.
(d) In this configuration, the ego-car is placed at intervals of 30 meters along the road. Raycasting is performed at 60-meter intervals. Visually, it looks mostly similar to the baseline. 18 yellow circles are overlaid on the map to mark changes compared to the baseline configuration (a). Most of them are located where the road transitions from adequate (green) to limited (red) visibility or vise versa, usually prolonging the red segments. Sometimes, a short red segment appears within a green segment.}
\end{figure*}

\new{We evaluate the robustness of our assistant's computation of available sight distance with respect to the granularity of the system’s raycasting intervals $d_r$ (baseline: 10m) and the sampling frequency of the ego-car’s position (baseline: 10m, i.e., compute the available sight distance every 10m).
On our virtual road network, the median available sight distance during overtakes was about 400m, which one requires at the point-of-no-return for overtaking (at 105km/h) a vehicle that drives 80km/h. 
Based on this scenario, we calculate the available sight distance for a variety of parameters (ego-car placement interval: $1-30$m, raycast interval $d_r$: $1-300$m). 
We illustrate our most important findings in Figure~\ref{fig:400m-sight-distance-raycast-egocar-params} and refer to the supplementary material for a complete list of plots.}

\new{Figure~\ref{fig:400m-sight-distance-raycast-egocar-params} displays our road network, indicating along the driving direction (black arrows in Figure~\ref{fig:400m-baseline}) where the required sight distance is available (green) and where it is not (red). 
Comparing our baseline (Figure~\ref{fig:400m-baseline}) to a finer resolution (Figure~\ref{fig:400m-1x1}), while a few red and green segments move by 10m, we see virtually no qualitative difference. 
One notable exception is in the center, where the baseline introduces a red dot within a green segment, causing a warning to briefly ($<1$s at 105km/h) appear. 
Comparing our baseline to less frequent raycasts ($d_r = 60m$, Figure~\ref{fig:400m-10x60}), we mostly see slightly longer red segments in the marked spots, prolonging the warning display. 
Most notably, in the lower middle, two small green segments are introduced within one red segment, which would cause a warning to briefly disappear. 
To a lesser extent, this also occurs with $d_r = 30m$. 
Placing the ego-car less frequently (every 30m) with $d_r = 60m$ (Figure~\ref{fig:400m-30x60}) had similar effects. 
Even less frequent raycasts ($d_r >= 120m$) introduced noticeable qualitative differences regardless of ego-car placement interval (supplementary material), increasingly undermining the assistant's ability to appropriately warn drivers.
While we only report one scenario for brevity, we have observed comparable effects for lower sight distances, with the baseline performing well.}

\section{Discussion}
\subsection{\ref{item:rq-acceptance}: How do drivers perceive and accept our assistant (in terms of workload and usability) across the UIs?} %
We find in Section~\ref{subsec:assis-yay-or-nay} that all but one participant preferred using our assistant over no assistant, with some even expressing they missed it.
Many appreciated how it eased the decision to overtake, reducing mental load and making them feel more relaxed (Section~\ref{subsec:indiv-ui-prefs}), 
The majority would almost never turn off the assistant with either UI, using it permanently or on unfamiliar country roads (Sections~\ref{subsec:sus} and~\ref{subsec:indiv-ui-prefs}).
In total, this matches our expectation of the majority favoring the assistant to driving without, 
contrasting Hegeman et al.~\cite{Hegeman2007}, where users rated the usefulness of a comparable overtaking assistant as low.
\new{This lays the groundwork for future studies to further validate the assistant's (perceived) usefulness. 
While we found the majority of participants would reportedly almost never turn off the assistant, our study did not focus on the correct perception of the warning and whether the timing felt appropriate. 
In future studies and with our improved prediction models, participants could be asked to actively indicate (e.g., via button press) when they would expect or accept a warning based on the driving context, and why. 
In addition to eye-tracking, driver reactions such as braking and acceleration patterns in response to the warning and reaction times could be tracked. 
}

While drivers liked the idea of the assistant, they were divided on the best UI. 
Although the monitoring-focused UI had a significantly higher System Usability score (Section~\ref{subsec:sus}), supporting~\ref{item:rq1-hyp-monitoring-higher-sus}, we learned in Section~\ref{subsec:indiv-ui-prefs} that the preferred UI heavily depends on personal preference and relates to situational use (scheduling-focused UI) vs permanent use (monitoring-focused UI).
\new{Anecdotally,} a car dealer noted that mid-age and older clients (who are more likely to buy more modern cars) likely prefer the simpler monitoring-focused UI due to difficulties with complex driving assistants
\new{%
-- verifying this requires a tailored user study.}
\new{A few participants challenged the system, hinting a desire for explanations.} %
\new{Since only few asked for it, adding icons for curves, hills, and intersections could be an optional feature.
While the transparency of our model eases explanations, in the spirit of Section~\ref{subsec:prediction-model} we would consult literature (on explainable artificial intelligence (XAI) in the automotive context) to underpin our design rationale before implementation.}

Regarding the DALI global score for perceived workload, no significant differences were found between conditions, offering no support for~\ref{item:rq1-hyp-no-higher-workload}. 
While some participants valued the scheduling-focused UI's ability to find overtaking opportunities, 
some reported it drew increased visual demand and attention, as reflected in the respective DALI categories (Section~\ref{subsec:dali}). 
Concerning the audio demand, we conjecture that, since no audio was involved besides driving/airflow and braking sound, the increased visual demand and attention of the scheduling-focused UI might have suppressed and thus lowered the perceived auditory "demand".
Furthermore, some drivers found the scheduling-focused UI's values inconsistent and deemed the update rate of 1Hz too fast.
For example, an overtaking opportunity could disappear if the vehicle to be overtaken accelerated too much.
We learned that already three elements can be overwhelming, and while we cannot compare our results to Hassein et al.~\cite{Hassein2019} due to the lack of a user study, we see similarities to Steinberger et al.~\cite{Steinberger2016} in terms of attention-demanding elements.

\subsection{\ref{item:rq-safety}: Which UI helps drivers (more) to avoid dangerous overtaking maneuvers?}\label{subsec:disc-rq-safety}
First and foremost, we find in Section~\ref{subsec:driving-data} that with both UIs, drivers follow a vehicle significantly longer before over\-taking it, with no significant differences between the UIs. 
This supports~\ref{item:rq2-hyp-follow-longer} and shows that either UI facilitates a more patient driving style. 
\new{
The observed delay in overtaking with the assistant may reflect the additional information drivers process to make informed decisions. 
This aligns with the assistant’s purpose to prioritize deliberate and cautious overtaking decisions over hastiness. 
Notably, the monitoring-focused UI required the least attention (Section~\ref{subsec:dali}), suggesting that it does not overload the driver with excessive cognitive demands but instead supports safer situational awareness and decision-making.}
\new{Participants reflect this in their interviews (Sections~\ref{subsec:dali} and~\ref{subsec:indiv-ui-prefs}). 
While they felt the overtaking maneuver itself is not made safer by the assistant, which is reinforced by us not finding qualitative differences in the overtaking maneuvers themselves between conditions (Section~\ref{subsec:driving-data}), it reportedly does (in their preferred UI) make the \textit{overall driving} safer by enabling a more patient driving style and by facilitating the decision of whether to overtake.}
\new{Overall, this indicates that our assistant, while not affecting \textit{how} drivers overtake, succeeds in enhancing decision-making regarding \textit{when} to overtake.}

Contrary to our expectations, no significant difference was found between conditions regarding the ratio of available to required sight distance at the measured point-of-no-return, providing no support for~\ref{item:rq2-hyp-increase-sight-distance}. 
While ratios below 1 were rare, ensuring ratios $\geq1$ would increase overtaking safety. 
\new{In this context, we found that the assistant (regardless of UI) sometimes mispredicted the point-of-no-return. %
While we emphasize that the measured point-of-no-return is based on the actual driving data rather than the assistant's prediction, we note that prediction errors could have indirectly influenced this data by affecting drivers' decision-making. %
For instance, a driver could have initiated the overtaking maneuver when no warning was shown (and no oncoming traffic was visible), and then reached an earlier-than-predicted point-of-no-return (because of higher-than-assumed speed), where there was less sight distance than required. 
While this does not change our interpretation of the results in Section~\ref{subsec:driving-data} -- in its current form, the assistant does not increase overtaking safety in terms of the available/required sight distance ratio -- we suspect this might change with correct predictions.
To this end, our study and running multiple acceleration models in parallel allowed us to identify improvements to near-zero relative error regarding the point-of-no-return prediction, opening the door to verify our expectation in a follow-up study. 
}

\new{With respect to the UIs, participants} found the monitoring-focused UI significantly less distracting than the scheduling-focused UI (Section~\ref{subsec:dali}), supporting~\ref{item:rq2-hyp-monitoring-less-distracting}. 
Similarly, the monitoring-focused UI takes perceived load off drivers in the DALI categories "Attention" and "Visual Demand". 
In connection to the findings on~\ref{item:rq-acceptance}, we conclude that while the monitoring-focused UI likely is the better default for many, both UIs have value, subject to improvements to the scheduling-focused UI. 
First, reduce inconsistencies in distance estimates by, e.g., filtering out short overtaking windows, adjusting the update rate, and rounding distance display (e.g., to the nearest 50m).
Second, create a third UI with the same features as the scheduling-focused UI but without the red distance estimate for users who find it distracting.
Furthermore, both UIs might profit from (optionally) showing the reason for warnings.

\subsection{\ref{item:rq-models}: %
How accurate are the sight distance prediction models used by our assistant?}
By combining technical contributions with user research, we gain insights that would be difficult to obtain if the two were separated like in, e.g., \cite{Hassein2019}.
Through a collection of the many user-performed overtaking maneuvers, we are able to detect modeling flaws and to adapt our model accordingly with our model understanding.

We find that even the simplest \textit{Constant} acceleration model, if tuned properly, proved effective for our assistant, providing no support for~\ref{item:rq3-hyp-ldm-and-dynamic-are-better}.
While the \textit{Dynamic} model does suggest potential with its many parameters, it is questionable whether tuning them further is worth the effort, as it is also computationally more expensive. 
Although we calculated predictions using all three models, warnings were based on the simplest one (Section~\ref{subsec:environment}).
Participants did not seem to notice that, depending on their driving style, the predictions of the point-of-no-return and consequently, the warnings %
were off at times. 
While having many more drivers engaging with the system even longer might make this more noticeable (this does call for further research), 
one possible explanation is that %
the sight distance at the measured point-of-no-return was usually sufficient (Section~\ref{subsec:driving-data}).

The prediction mismatch became clear only after analyzing the driving data, i.e., the results in Section~\ref{subsec:driving-data} motivated the %
questioning of key model assumptions in Section~\ref{subsec:model-comparison}. 
This in turn yielded valuable insights into how the parameters can be adapted -- an emphasis on how intertwined technical aspects are with user behavior.

\new{Our most important learning after parameter adaptation is that our models are indeed able to match the predicted point-of-no-return with the measured one from the driving data -- at a cost.
More precisely,}
we learn that, to improve predictions, we should change the acceleration coefficient~$\lambda$ from $0.8$ to $1$.
This \new{"cheap" or trouble-free adjustment} better reflects drivers completing such overtaking maneuvers as quickly as possible. 
We also learn that a huge improvement comes not from the acceleration model, but 
from lifting the maximum velocity constraint.
\new{We deem this a more "expensive" or problematic adjustment as} this risks encouraging traffic violations.
Legally, one way out could be to put all responsibility on the driver and lift the constraint.
Arguably, if combined with other assistants, such as those assisting with acceleration or braking, lifting the constraint might help in emergencies, e.g., if finishing the maneuver faster avoids a collision with opposing traffic.
However, our assistant is meant to \textit{warn} of dangerous overtaking maneuvers to enhance driving safety.
Encouraging speeding goes against that spirit. 
Moreover, our warning is issued while drivers follow a vehicle in front, giving the driver time to observe traffic and specifically avoid such emergency situations. 
Hence, \new{although the easiest route to follow-up studies on available/required sight distance ratios from Section~\ref{subsec:disc-rq-safety},} we do have reservations about lifting the maximum velocity constraint %
\new{in a real-world implementation}. %

A possible research avenue in particular for the monitoring-focused UI is to predict not one point-of-no-return but several, including those that can be reached by speeding, and \textit{not} show a warning only if the sight distance is sufficient at all the predicted points. 
Future work could investigate the created trade-off between enhancing safety in particular for speeding drivers and potentially fewer overtaking opportunities in particular for law-abiding drivers. 

\new{Beyond that, future work could examine the sight distance prediction models in different conditions, such as weather (rain, snow, fog) or low-light conditions, which affect driving, vehicle behavior, and sight distance. 
To better gauge model performance in the real world, one could investigate what adjustments need to be made for these conditions (in particular regarding the key model assumptions from Section~\ref{subsec:model-comparison}), and whether the \textit{Dynamic} model can leverage its many parameters to perform well under various conditions.} %

\subsection{Treatment of World Knowledge}
We designed our driving assistant to rely solely on 3D map data and today's in-car sensors like LiDAR, radar, and cameras (Section~\ref{subsec:sensors-available}).
To speed up prototyping and testing, we used a simulation environment, accessing vehicles' speed and length directly. 
To simulate real-world sensor constraints, we mimicked the range and limited the assistant’s access to information, only considering vehicles within realistic sensor range. 
We also limited the update rate to 1Hz.
This methodological convenience let us focus on our assistant's decision-making rather than simulating raw sensor data.
While not an indication of reliance on world knowledge, 
this approach clearly has limitations.

We idealized the simulation by avoiding raw sensor data and excluding noise and latency found in real sensors. 
While this allowed us to better isolate how our acceleration models affect safe sight prediction in Section~\ref {subsec:model-comparison}, this also meant our assistant knew precisely the length $L_{oen}$ and speed $v_{oen}$ of the vehicle to be overtaken. %
In real-world use, adaptive cruise control (ACC) relies on speed measurements; the same technology can be used in our assistant. 
The length $L_{oen}$ could be established by combining camera-based visual cues that detect the vehicle's endpoints with radar or LiDAR, which measure the distance to the front and rear of the vehicle. 
Measurement tolerances could be incorporated since an upper estimate of $L_{oen}$ suffices. 
While overestimating $L_{oen}$ might lead to slightly more/longer warnings, it would not compromise safety.
Future work could integrate more sophisticated sensor simulations with noise, latency, and data loss. 
To this end, Wang et al.~\cite{wang2023robust} recently proposed a promising extended Kalman filter.

A more critical role in the safe sight distance prediction %
is played by the temporal components.
Both the safety distance (in seconds) between the overtaking and overtaken vehicle \textit{after} the maneuver, $L_2^s$, and the safety distance (in seconds) between the overtaking vehicle and oncoming traffic after the maneuver, $L_{sm}^s$, were carefully set according to literature (Section~\ref{subsec:calculating-safety-distances}).
Driving data supports the chosen value for $L_{sm}^s$. 
The larger-than-expected $L_2^s$ observed (i.e., drivers were more cautious when reeving than anticipated) requires further research, as it might be due to participants' difficulties in estimating the distance to rear traffic through the mirrors in the VR simulation. 
Ultimately, both parameters could be user-adjustable within reasonable ranges (1-2s for $L_2^s$, 1.5-2s for $L_{sm}^s$), like the distance setting in ACC assistants.

Another simplification is that detecting two vehicles ahead and calculating their distance always worked perfectly within sensor range. 
This let us assess preferences for an additional warning icon for too little distance between two vehicles ahead, which two participants liked. 
We argue that modern sensors like cameras, radar, and LiDAR can often infer a second vehicle’s presence, even without seeing through objects. 
On straight roads, both driver and sensors can detect it in phase (i). 
Around curves or hills, cameras may spot parts of the second vehicle such as its top or its lights, while radar can detect object edges corresponding to a second vehicle.
Deceleration patterns of the vehicle directly ahead could serve as additional cues. 
Image processing could then estimate the second vehicle's distance, $L_e$, based on its apparent size and position. 
However, while the sensors exist, real-world use would require tailored algorithms for these (fused) sensors and reliability testing.
Recent advances include the works of Hess et al.~\cite{hess2024lidarclip} and Ljungbergh et al.~\cite{ljungbergh2023raw}, with pending patents such as~\cite{patent2023:predicting-behavior}.
We can facilitate this process to some extent, as the exact $L_e$ is not necessary -- lower estimates can ensure safety. 
Finally, if the system recognizes two vehicles but cannot estimate $L_e$, the driver could be warned of \textit{potentially} too little distance.

A key uncertainty is the presence and speed of oncoming traffic, which current sensor ranges cannot detect in time. 
Therefore, we do not use simulation data for oncoming traffic and instead always assume its presence for safety. 
Based on empirical data from literature, we developed the model~\eqref{eq:vmaxopp} for its speed $v_{opp}^{max}$, which linearly depends on the speed limit. 
A 100km/h speed limit results in $v_{opp}^{max}\approx 124$km/h, which we can easily adjust with more driving data.
While we do not explicitly anticipate more excessive speeding of oncoming traffic on (curvy) rural roads, this model captures cases where oncoming traffic sees the ego-car overtaking and reduces their speed accordingly. 
Moreover, given participants' interest in an assistant that accounts for oncoming traffic, future work could include detecting and warning about oncoming vehicles within sensor range.

Overall, we believe this design maintains the assistant's independence from a priori world knowledge besides available 3D map data, thus distinguishing it from systems that rely on detailed external information.
Our precise model description and the recorded driving data helps establishing and validating a sensor processing pipeline in future work.

\subsection{Other Questions and Limitations}

Our assistant relies on the availability of DSM data, obtainable from, e.g., \url{https://www.opendem.info/opendemsearcher.html} by selecting a region. 
\new{We found that data with a grid spacing of 1m or 2m is available.}
If no or only coarse DSM is available in the current region, we follow the common procedure %
of informing the driver that the assistant is currently unavailable~\cite{web:tesla-troubleshooting}.
\new{While a proper quantification of what would be "too coarse" requires assessing the assistant's behavior on real-world sample data, our analysis in Section~\ref{subsec:robustness} provides some first impressions. 
In it, we find that the calculation of available sight distance is quite robust: even raycasts every 60m cause only minor deviations compared to our baseline (10m ego-car placement interval, 10m raycast interval). 
Since the finer resolution (both 1m) improved results only marginally, we are confident that (after potential fine-tuning, see below) our assistant works with real-world data, even at slightly lower resolution than a 1m grid spacing.}

\new{The main effect of less frequent raycasts~$d_r$ and ego-car positions (Section~\ref{subsec:robustness}) was slightly prolonged red sections (and thus warnings). 
While not immediately dangerous, drivers might eventually notice inconsistencies and question the assistant’s reliability, which is unlikely with our baseline settings. 
Another issue was very short green segments, where a brief warning disappearance could confuse drivers. 
To prevent this, short green segments (e.g., $<$50m) could be ignored, keeping the warning active. 
Similarly, short red segments occasionally appeared between green sections, but for raycast intervals~$d_r\leq60$m, these only followed a red segment, making it simple to prolong the warning. 
Red segments in the middle of large green sections occurred only for $d_r>60$m, suggesting these intervals are unsuitable.}

In our experiment, we pre-programmed the route in the navigation system. 
In practice, if no destination is set, a most probable path algorithm~\cite{mppPatent} can be used to predict the driver's route, albeit with uncertainty.

A potential issue in real-world use is the risk of misinterpreting the assistant. 
In our study, we clearly told participants that no warning does not mean a suggestion to overtake, as oncoming traffic is not detected, and drivers are responsible for their actions. 
\new{While none of the study participants suggested a UI change in this regard and the disappearance of a warning in either UI did not trigger an immediate initiation of the overtaking maneuver (Section~\ref{subsec:driving-data}),}
in practice, users must be informed about the system’s limits\new{, e.g.,} via manuals, sales staff, or in-car prompts. 
\new{Future studies could shed light on how to best communicate these limits, whether the absence of a warning leads to drivers misinterpreting the assistant, and if so, how to combat it.
This is particularly important if one can deactivate the assistant, since the absence of a warning could then have two causes (deactivation or sufficient sight distance).}
For example, when activating the assistant, a reminder could inform drivers that no warning does not imply safety or an endorsement to overtake, similar to driving backwards with a camera. 
\new{Beyond that}, driver behavior could be monitored to inform the driver, e.g., if the assumed safety distance thresholds are violated such that it affects prediction reliability. %

Another under-explored problem in our work is the real-world implementation of the HUD. 
In our simplified design, speedometer and rpm gauge were displayed on a standard dashboard interface. 
The HUD was dedicated to the assistant and did not display other information such as the maximum allowed speed or navigation instructions. 
Our approach allowed us to clearly convey the assistant's core functionalities, ensure the UI elements are always visible in the limited field-of-view of the VR headset, and evaluate our initial design in terms of how it is perceived and whether and how it helps drivers to avoid dangerous overtaking maneuvers. 
In real-world use, however, the assistant's information might have to compete for the limited HUD space. 
For studies that increase realism by integrating the assistant's warnings into existing HUDs and balancing the display of overtaking-related warnings with other information, this work could serve as a baseline. 
For example, if drivers had to choose bet ween information layers on the HUD, how would this choice affect their decision-making regarding overtaking? 
In this context, future studies could also revisit the chosen symbols, analyzing their intuitiveness and integrability.

Our research focused on improving drivers' decision-making \textit{before} overtaking, while behavioral metrics \textit{during} overtaking maneuvers were assessed qualitatively. 
Concerning these metrics, we found no clear differences between conditions, apart from a possible trend toward more safety distance to opposing traffic with the monitoring-focused UI.
While a statistical analysis could add robustness, potential inaccuracies in the point-of-no-return prediction might have influenced this data, complicating reliable conclusions. 
Future research could address this by refining the dataset or collecting new data with improved prediction accuracy.

Lastly, assuming all cars drive autonomously before our assistant's market launch, the safe sight prediction and a similar UI could be used to explain to the driver/passenger, e.g., why the vehicle does not overtake even if instructed to.

\section{Conclusion}
In this paper we proposed an overtaking assistant that warns users of dangerous overtaking maneuvers on country roads. 
We excluded vehicle-to-vehicle communication and limited ourselves to the range of current in-car sensors, which necessitated the new model for sight distance prediction presented in this paper.
We implemented a prototype of the assistant with two UI variants -- one focused on monitoring, the other on scheduling -- in Unity and evaluated it in a user study with 25 participants.
Our results show that participants welcome the assistant, which enables more patient driving with both UI variants.
While the less distracting monitoring-focused UI achieves a higher System Usability Score, the preferred UI depends on personal preference and relates to situational (scheduling-focused) vs permanent (monitoring-focused) use. 
Through combination of technical contributions with user research we identified crucial model parameters and assumptions that significantly improved model predictions.
We are confident that the ideas and models in this paper can forward the design of future warning assistants. 
To this end, we open-source TOWARDS, The Overtaking Warning Assistant Recorded Data Set, and our safe sight prediction models.

\begin{acks}
We express our sincere thanks to Professor Jörg Müller from the Chair of Serious Games for providing the simulator hardware and valuable feedback. 
We also extend our thanks to Professor Daniel Buschek for the many fruitful discussions, which greatly helped to improve our work.
Arthur Fleig acknowledges the financial support by the Federal Ministry of Education and Research of Germany and by the Sächsische Staatsministerium für Wissenschaft, Kultur und Tourismus in the program Center of Excellence for AI-research „Center for Scalable Data Analytics and Artificial Intelligence Dresden/Leipzig“, project identification number: ScaDS.AI.
\end{acks}

\bibliographystyle{ACM-Reference-Format}
\bibliography{sample-base-camera-ready}

\appendix

\section{Details on the Dynamic Acceleration Model and on Model Parameters}\label{app:dyn-acc}
The dynamic acceleration model~\cite{Rakha2002,acceleration_models} is based on forces:
\begin{equation}
\hn{a} = \frac{\hn{F_A} - \hn{F_R}}{\hn{m}}
\end{equation}
Driving forces $F_A$ are limited by wheel slip and engine power:
\begin{equation}
\hn{F_A} = \min \left(\hn{\eta} \cdot \hn{\beta} \cdot \frac{\hn{P}}{\hn{v}}, \hn{m_t} \cdot \hn{g} \cdot \hn{\mu}\right)
\end{equation}
where 
\begin{flalign*}
\text{\hspace{\leftmargin}} \hn{\beta} &= \text{variable power factor} && \\
\hn{\eta} & = \text{transmission efficiency} && \\
\hn{P} & = \text{engine power} && \\
\hn{v} & = \text{car speed} && \\
\hn{m_t} & = \text{mass of vehicle on tractive axle} && \\
\hn{g} & = \text{gravitational constant} && \\
\hn{\mu} & = \text{coefficient of friction between tires and pavement} && \\
\end{flalign*}
Resistant forces $F_R$ include air resistance including opposing wind, wheel friction, and road incline:
\begin{equation}
\hn{F_R} = \frac{1}{2} \hn{A_f} \cdot \hn{c_w} \cdot \hn{\rho_{Air}} \cdot \left(\hn{v} + \hn{v_{wind}}\right)^2 + \hn{g} \cdot \hn{m} (1 - \hn{G}) \cdot \hn{C_r} + \hn{g} \cdot \hn{m} \cdot \hn{G}
\label{konzept:fahrzeug:dyn:rakha:fr}
\end{equation}
where
\begin{flalign*}
\text{\hspace{\leftmargin}} \hn{A_f} & = \text{vehicle frontal area} && \\
\hn{c_w} & = \text{vehicle drag coefficient} && \\
\hn{\rho_{Air}} & = \text{air density} && \\
\hn{C_r} & = \text{rolling coefficient} && \\
\hn{v_{wind}} & = \text{opposing wind speed} && \\
\hn{m} & = \text{car mass} && \\
\hn{G} & = \text{road incline in percent} && \\
\end{flalign*}
Some of these parameters can be directly extracted from the car data sheet, others have to be tuned. %

Table~\ref{tab:simulator:parameter} lists all model parameters used for the simulation for each of the acceleration models.
\begin{table}[ht]
\centering
\begin{tabular}{c | c | c } 
Model & Variable & Value \\ \hline
\multirow{10}{*}{Dynamic} & $\hn{m}$ & $2300 \text{ } kg$ \\
& $\hn{P}$ & $182.7 \text{ } kW$ \\
& $\hn{\mu}$ & $1.495$ \\
& $\hn{\eta}$ & $1$ \\
& $\hn{\beta}$ & $0.905$ \\
& $\hn{A_f}$ & $2.54 \text{ } m^2$ \\
& $\hn{c_w}$ & $0.381$ \\
& $\hn{\rho_{Air}}$ & $1.204 \text{ } kg/m^3$ \\
& $\hn{C_r}$ & $5.98 \cdot 10^{-12}$ \\
\hline
\multirow{2}{*}{LDM} & $\hn{a_{max}}$ & $7.80 \text{ } m/s^2$ \\
& $\hn{v_e}$ & $41.85 \text{ } m/s$ \\
\hline
Constant & $\hn{a_{const}}$ & $3.00 \text{ } m/s^2$ \\
\hline
GLOBAL & $\hn{\lambda}$ & $0.8$ \\
\end{tabular}
\caption{Model parameters used for the simulation for each of the acceleration models}
\label{tab:simulator:parameter}
\end{table}

\section{More Details on the Demographic Data of Participants}\label{app:participants}
\begin{figure}[ht]
\includegraphics[width=\linewidth]{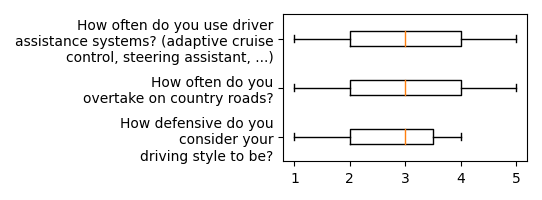}
\caption{Boxplot of the 5-point Likert scale analysis of questions related to driving \label{fig:demographic_likert}}
\Description{This figure shows three box plots with scale ranging from 1 to 5. The first question is, "How often do you use driver assistance systems?", with broad distribution from 1 to 5, median 3 and box ranging from 2 to 4. The second question is, "How often do you overtake on country roads?", with with visually the same results as the previous plot. The third question is, "How defensive do you consider your driving style to be?", with median 3 and box from 2 to 3.5.}
\end{figure}

\begin{figure}[ht]
\includegraphics[width=\linewidth]{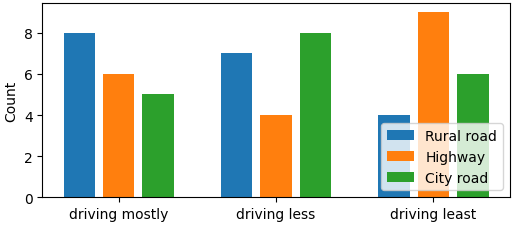}
\caption{Participants' responses to the question "On which road type do you drive the most?" \label{fig:demographic_driving}}
\Description{This figure shows a bar plot with "Count" on the y axis (values from 0 to 9), and three options "driving mostly", "driving less" and "driving least" on the x axis. For each of these three options, there are three bars: Blue for "Rural road" (mostly: 8; less: 7; least: 4), orange for "Highway" (mostly: 6; less: 4; least: 9) and green for "City road" (mostly: 5; less: 8; least: 6).}
\end{figure}

Three participants regularly transport kids in their car.
Ten participants regularly drive with their partner.
Other questions participants answered before driving are shown in Figures~\ref{fig:demographic_likert} and~\ref{fig:demographic_driving}.

\section{TOWARDS: The Overtaking Warning Assistant Recorded Data Set}\label{app:logdata}
The CSV files of \textbf{T}he \textbf{O}vertaking \textbf{W}arning \textbf{A}ssistant \textbf{R}ecorded \textbf{D}ata \textbf{S}et (TOWARDS), 6.5 GB in total, include:
\begin{itemize}
\item Time: Absolute time, simulation time without breaks
\item Driver's car: 3D position, 3D rotation, 3D velocity vector, 3D angular velocity vectors
\item Driver input: acceleration pedal input, break pedal input, steering angle
\item Of up to 4 vehicles in front of or overtaken by the driver: 3D position, distance to driver's car, vehicle length, speed, planned direction at the next intersection, unique ID
\item Next oncoming vehicle: 3D position, distance to driver's car, speed, length
\item Overtaking warning assistant values:
$L_1$, $L_{oen}$, $L_{oing}$, $L_2$, $L_{tot}$, $L_e$, $L_3$, $G$, $v^{max}_{opp}$, warning (yes/no), warning reason
\item For each acceleration model: $d_{rest}$, predicted total overtaking duration, predicted time and distance until point-of-no-return is reached, $v_{oing,end}$, $d_{opp}$, $L_{sm}$, $d_{s,min}$, distance to next overtaking opportunity
\item Number of lanes per direction on the current road, distance of the front left wheel to the center line of the road, distance traveled along the current road, total length of the current road
\item Visibility: Currently available sight distance
\item Collisions: Number of collisions, Number of severe collisions (Unity impulse > 100), 3D position of the collided object, 3D Unity impulse of the collision, Collision type (vehicle, terrain, unknown)
\item Manual button presses: Number of re-centerings of the VR headset, Number of manual resets to set the car back on the road, Number of severe collisions as deemed by the experimentor, Number of manual markings of events deemed special by the experimentor
\end{itemize}

\section{Results on Perceived Safety}\label{app:perceived-safety}
\begin{figure}[ht]
\includegraphics[width=\linewidth]{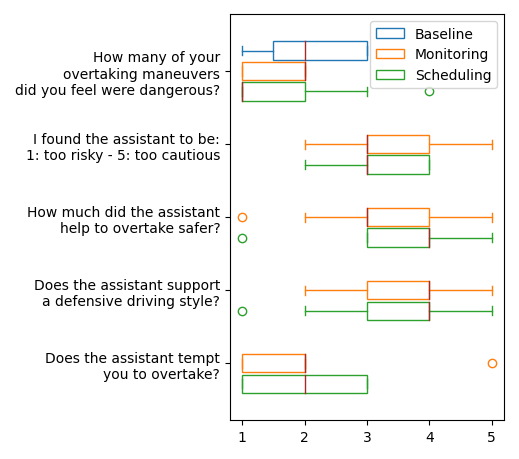}
\caption{Boxplot of the 5-point Likert scale analysis of questions related to perceived safety. No outliers are hidden by the legend. 
\label{fig:perceived-safety}}
\Description{This boxplot illustrates the boxplot of the 5-point Likert scale analysis (values 1-5) of five questions related to perceived safety. 
The first question is, "How many of your overtaking maneuvers did you feel were dangerous?". The "Baseline" condition exhibits values between 1-3, with a median of 2 and box ranging from 1.5 to 3. The "Monitoring" condition has values between 1-2, a median of 2, and box ranging from 1 to 2. The "Scheduling" condition exhibits values between 1-3, an outlier at 4, a median of 1, and box ranging from 1 to 2. 
The second question is, "I found the assistant to be 1:too risky - 5:too cautious". The "Monitoring" condition has values between 2 and 5, a median of 3, and box ranging from 3 to 4. The "Scheduling" condition exhibits values between 2 and 4, a median of 3, and box ranging from 3 to 4. 
The third question is, "How much did the assistant help to overtake safer?". The "Monitoring" condition has values between 2 and 5, an outlier at 1, a median of 3, and box ranging from 3 to 4. The "Scheduling" condition exhibits values between 3 and 5, an outlier at 1, a median of 4, and box ranging from 3 to 4. 
The fourth question is, "Does the assistant support a defensive driving style?". The "Monitoring" condition has values between 2 and 5, a median of 4, and box ranging from 3 to 4. The "Scheduling" condition looks the same except one outlier at 1. 
The fifth question is, "Does the assistant tempt you to overtake?". The "Monitoring" condition has values between 1 and 2, an outlier at 5, a median of 2, and box ranging from 1 to 2. The "Scheduling" condition exhibits values between 1 and 3, a median of 2, and box ranging from 1 to 3.}
\end{figure}

The questions related to perceived safety and participants' responses are illustrated in Figure~\ref{fig:perceived-safety} for the sake of completeness. 
\new{For the first question regarding the number of dangerous overtaking maneuvers, 1 corresponds to 0\%, and 5 to 100\%. For the last three questions, 1 corresponds to "not at all", and 5 to "very".}
Most notably, both UI variants are deemed on the safer side of cautiousness, though not too cautious. 
No significant differences between conditions were found in the five questions.

\section{Effect sizes of model comparisons}\label{app:model-comp-effect-sizes}
Table~\ref{tab:app:eq_time_data} shows the resulting effect sizes of the three compared models and their tweaked variants. All reported significance levels are $p<0.001$ using Wilcoxon method except for the comparison between constant and dynamic model for $\lambda=1$ case, where $p=0.02$.
\begin{table*}[ht]
\centering
\begin{tabular}{l || c | c | c | c | c | c | c | c} 
& \multicolumn{2}{c|}{Constant} & \multicolumn{3}{|c|}{LDM} & \multicolumn{3}{|c}{Dynamic} \\
& $\lambda = 1$ & w/o limits & as modeled & $\lambda = 1$ & w/o limits & as modeled & $\lambda = 1$ & w/o limits \\ \hline \hline
\multicolumn{8}{l}{Constant} \\ \hline
\makecell[l]{as\\modeled}&\makecell{Glass's\\$\Delta=0.156$\\Effect:\\Very small}&\makecell{Glass's\\$\Delta=2.405$\\Effect:\\Huge}&\makecell{Cohen's\\$d=0.028$\\Effect:\\Very small}&-&-&\makecell{Cohen's\\$d=0.061$\\Effect:\\Very small}&-&- \\ \hline
$\lambda = 1$&-&\makecell{Glass's\\$\Delta=1.973$\\Effect:\\Very large}&-&\makecell{Cohen's\\$d=0.018$\\Effect:\\Very small}&-&-&\makecell{Wilcoxon\\$p=0.02$\\Cohen's\\$d=0.003$\\Effect:\\Very small}&- \\  \hline
\makecell[l]{w/o\\limits}&-&-&-&-&\makecell{Cohen's\\$d=0.440$\\Effect:\\Small}&-&-&\makecell{Cohen's\\$d=0.121$\\Effect:\\Very small} \\ \hline

\multicolumn{8}{l}{LDM} \\ \hline
\makecell[l]{as\\modeled}&-&-&-&\makecell{Cohen's\\$d=0.157$\\Effect:\\Very small}&\makecell{Glass's\\$\Delta=1.628$\\Effect:\\Very large}&\makecell{Cohen's\\$d=0.035$\\Effect:\\Very small}&-&- \\ \hline
$\lambda = 1$&-&-&-&-&\makecell{Glass's\\$\Delta=1.240$\\Effect:\\Very large}&-&\makecell{Cohen's\\$d=0.021$\\Effect:\\Very small}&- \\ \hline
\makecell[l]{w/o\\limits}&-&-&-&-&-&-&-&\makecell{Cohen's\\$d=0.316$\\Effect:\\Small} \\ \hline

\multicolumn{8}{l}{Dynamic} \\ \hline
\makecell[l]{as\\modeled}&-&-&-&-&-&-&\makecell{Cohen's\\$d=0.211$\\Effect:\\Small}&\makecell{Glass's\\$\Delta=2.293$\\Effect:\\Huge} \\ \hline
$\lambda = 1$&-&-&-&-&-&-&-&\makecell{Glass's\\$\Delta=1.692$\\Effect:\\Very large} \\ \hline

\end{tabular}
\caption{Effect sizes of model comparisons}
\label{tab:app:eq_time_data}
\end{table*}

\section{\new{Behavioral metrics during overtaking}}\label{app:qual-overtaking}
\begin{figure*}[ht]
\centering
\subcaptionbox{\new{Maximum speed while overtaking (in km/h)}\label{fig:safety-app-1}}%
[.49\textwidth]{\includegraphics[width=0.49\textwidth]{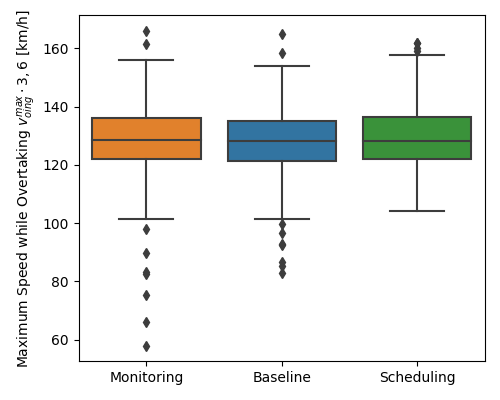}}
\subcaptionbox{\new{Duration on the left lane (in seconds)}\label{fig:safety-app-2}}%
[.49\textwidth]{\includegraphics[width=0.49\textwidth]{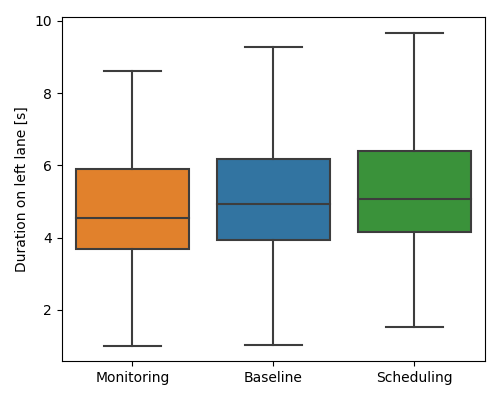}}
\subcaptionbox{\new{Rearward safety distance $L_2^s$ (in seconds)}\label{fig:safety-app-3}}%
[.49\textwidth]{\includegraphics[width=0.49\textwidth]{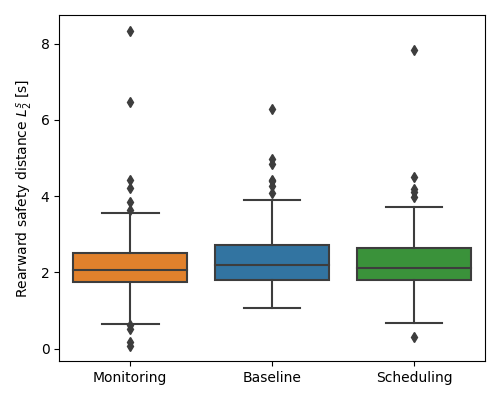}}
\subcaptionbox{\new{Safety distance to opposing traffic $L^s_{sm}$ (in seconds)}\label{fig:safety-app-4}}%
[.49\textwidth]{\includegraphics[width=0.49\textwidth]{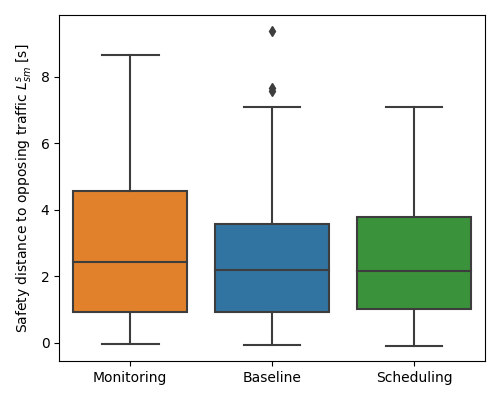}}
\caption{\new{Behavioral metrics during overtaking maneuvers}}
\label{fig:safety-app}
\Description{(a) Maximum speed while overtaking (in km/h)
The figure is a boxplot illustrating the maximum speed while overtaking (measured in km/h) during overtaking maneuvers. It compares three conditions: Monitoring, Baseline, and Scheduling.
Axes: The y-axis is labeled "Maximum speed while overtaking $v_{oing}^{max}*3.6$ [km/h]" and ranges from 60 to 160 km/h, with increments of 20 km/h. The x-axis shows the three conditions: Monitoring, Baseline, and Scheduling.
Boxplot Details: Each condition is represented by a vertical boxplot. The boxes for the three conditions are color-coded: orange (Monitoring), blue (Baseline), and green (Scheduling). The horizontal line inside the box represents the median maximum speed. Outliers (data points outside the whiskers) are displayed as individual black dots.
Observations: The medians across the three conditions are visually similar, roughly centered around 120-140 km/h. There are noticeable outliers in all conditions: The "Monitoring" and "Baseline" conditions have several low outliers (below 100 km/h). The "Scheduling" condition has a few high outliers near 160 km/h. The spread of data is similar between conditions.
(b) Duration on the left lane (in seconds)
The figure is a boxplot illustrating the behavioral metric duration on the left lane (measured in seconds) during overtaking maneuvers. It compares three conditions: Monitoring, Baseline, and Scheduling.
Axes: The y-axis is labeled "Duration on left lane [s]" and ranges from roughly 1 to 10 seconds. The x-axis shows the three conditions: Monitoring, Baseline, and Scheduling.
Boxplot Details: Each condition is represented by a vertical boxplot. The boxes for the three conditions are color-coded: orange (Monitoring), blue (Baseline), and green (Scheduling). The horizontal line inside the box represents the median duration. No outliers are visible.
Observations: The medians across the three conditions are visually similar, centered around 5-6 seconds. The spread of durations, indicated by the box and whisker lengths, appears consistent across conditions, with no large deviations or skewness.
(c) Rearward safety distance after reeving (in seconds)
The figure is a boxplot illustrating the behavioral metric rearward safety distance after reeving (measured in seconds) during overtaking maneuvers. It compares three conditions: Monitoring, Baseline, and Scheduling.
Axes: The y-axis is labeled "Rearward safety distance $L_2^s$ [s]" and ranges from 0 to 8 seconds, with increments of 1 second. The x-axis shows the three conditions: Monitoring, Baseline, and Scheduling.
Boxplot Details: Each condition is represented by a vertical boxplot. The boxes for the three conditions are color-coded: orange (Monitoring), blue (Baseline), and green (Scheduling). The horizontal line inside the box represents the median rearward safety distance. Outliers (data points outside the whiskers) are represented by individual black dots.
Observations: The medians across the three conditions are visually similar, centered around 2-3 seconds. All three conditions display noticeable outliers: The "Monitoring" condition includes a few high outliers above 6 seconds and a few close to 0 seconds. The "Baseline" condition shows several high outliers, with one reaching approximately 6 seconds. The "Scheduling" condition has a high outlier near 8 seconds and a low outlier close to 0 seconds. The spread of data is comparable across conditions, with no drastic differences in interquartile ranges.
(d) Safety distance to opposing traffic (in seconds)
The figure is a boxplot illustrating the behavioral metric safety distance to opposing traffic after reeving (measured in seconds). It compares three conditions: Monitoring, Baseline, and Scheduling.
Axes: The y-axis is labeled "Safety distance to opposing traffic $L^s_{sm}$ [s]" and ranges from 0 to 8 seconds, with increments of 2 seconds. The x-axis shows the three conditions: Monitoring, Baseline, and Scheduling.
Boxplot Details: Each condition is represented by a vertical boxplot. The boxes for the three conditions are color-coded: orange (Monitoring), blue (Baseline), and green (Scheduling). The horizontal line inside the box represents the median safety distance. Outliers (data points outside the whiskers) are represented as individual black dots.
Observations: The medians across the three conditions are centered around 2-3 seconds. There are some outliers in the "Baseline" condition, with a few high values reaching above 6 seconds. No apparent outliers are seen in the "Monitoring" and "Scheduling" conditions. The spread of data (indicated by the interquartile range and whisker lengths) is relatively consistent across conditions, with "Monitoring" showing a slightly wider interquartile range compared to the others.}
\end{figure*}

\new{Figure~\ref{fig:safety-app} depicts behavioral metrics during overtaking maneuvers, with each subplot showing a key parameter: (i) top speed [km/h], (ii) duration in the overtaking lane [s], (iii) rearward safety distance after reeving [s], and (iv) safety distance to opposing traffic after the maneuver [s].%
}

\end{document}